\documentclass[aps,pra,superscriptaddress,twocolumn,longbibliography]{revtex4-2}

\usepackage[english]{babel}
\usepackage{amssymb}
\usepackage{algorithm}
\usepackage{algpseudocode}
\usepackage{graphicx}
\usepackage{color}
\usepackage[colorlinks=true,
			urlcolor=blue,
			citecolor=blue,
			linkcolor=blue,
			citebordercolor={1 0 0},
			linkbordercolor={0 0 1}]{hyperref}
\usepackage{quantikz}
\usepackage{soul}
\usepackage{xurl}

\usepackage{nameref}
\newcommand{\figref}[1]{Fig.~\ref{#1}}

\algrenewcommand\algorithmicrequire{\textbf{Input:}}
\algrenewcommand\algorithmicensure{\textbf{Output:}}

\newcommand{\bea}{\begin{eqnarray}}
\newcommand{\eea}{\end{eqnarray}}
\newcommand{\be}{\begin{equation}}
\newcommand{\ee}{\end{equation}}
\newcommand{\ba}{\begin{equation}\begin{aligned}}
\newcommand{\ea}{\end{aligned}\end{equation}}

\def\1{\mathds{1}}

\makeatletter

\usepackage{circuitikz}
\usepackage{ulem}
\begin{document}

\title{Quantum Resilience: Canadian Innovations in Quantum Error Correction and Quantum Error Mitigation}

\author{Gaurav Saxena}
\email{gaurav.saxena@lge.com}
\affiliation{LG Electronics Toronto AI Lab, Toronto, Ontario M5V 1M3, Canada}

\author{Jack S. Baker}
\email{jack.baker@lge.com}
\affiliation{LG Electronics Toronto AI Lab, Toronto, Ontario M5V 1M3, Canada}

\author{Pablo D\'iez Valle}
\affiliation{LG Electronics Toronto AI Lab, Toronto, Ontario M5V 1M3, Canada} 
\affiliation{Instituto Tecnológico de Galicia, Cantón Grande 9, Planta 3, 15003 A Coruña, Spain}

\author{William E. Salazar}
\affiliation{LG Electronics Toronto AI Lab, Toronto, Ontario M5V 1M3, Canada} 

\author{Kevin Ferreira}
\affiliation{LG Electronics Toronto AI Lab, Toronto, Ontario M5V 1M3, Canada} 

\author{Thi Ha Kyaw} 
\email{thiha.kyaw@lge.com}
\affiliation{LG Electronics Toronto AI Lab, Toronto, Ontario M5V 1M3, Canada}

\date{\today}

\begin{abstract}
In celebration of the 2025 International Year of Quantum Science and Technology, this article highlights the pioneering achievements and ongoing innovations in quantum error correction and quantum error mitigation by Canadian institutions, academia and industry alike. Emphasizing Canada's central role in advancing these two related areas, we summarize landmark theoretical breakthroughs, cutting-edge experiments, and emerging techniques aimed at reducing and/or eliminating errors incurred when using a quantum computer. This community-focused overview underscores Canada's leadership in addressing the critical challenge of noise in quantum information science.
\end{abstract}

\maketitle


Quantum computers promise to solve problems that are intractable even for the fastest supercomputers, from designing new materials to cracking complex optimization problems. Researchers across
the globe are tirelessly developing quantum solutions to tackle complex problems, and Canada is at the forefront of
these efforts. Notably, Canada is home to both the first quantum hardware company (D-Wave) and the first quantum
software company (1QBit), in the world.
Xanadu is another leading Canadian photonic quantum computing company that specializes in both hardware and software development. 
And, of course, there are many other startups and larger companies in Canada and worldwide that we do not have enough space to list them all here.
Despite these tremendous concerted efforts, significant challenges still stand in the way of deploying quantum
computers for real-world applications.  

\begin{figure*}[t]
        \centering
        \includegraphics[width=0.8\linewidth]{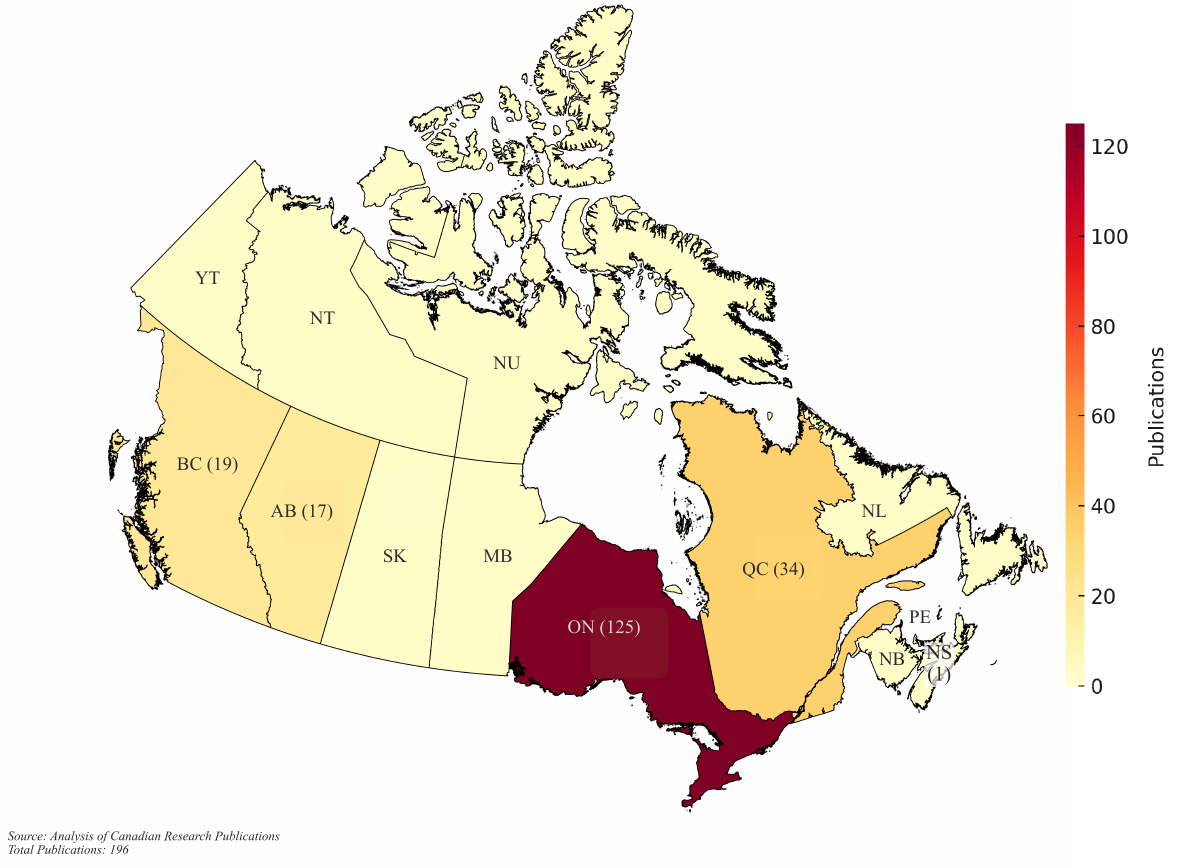}
        \caption{Geographical distribution of Quantum Error Correction and Quantum Error Mitigation papers, based on Canadian affiliated academic and industry researchers, scanned through the period of year 1990 to 2025, based on the Web of Science database.
    }
        \label{fig:Canada_heatmap}
\end{figure*}

One such challenge is the extreme fragility of quantum bits (qubits).
The slightest interaction with the environment or tiny imperfections in hardware can introduce errors that spoil a calculation, complicating both information storage and the scaling of quantum processing units. 
Even cosmic rays can deter the fate of qubits~\cite{Martinis2021Jun}!
Overcoming this ``noise” is essential to unlock useful quantum computing \cite{BibEntry2023Feb}. Researchers worldwide, including many in Canada, are developing two complementary strategies to tackle errors in quantum information processing. One approach is quantum error correction (QEC): encoding information in clever ways so that errors can be detected and fixed on the fly \cite{BibEntry2013Sep,Roffe2019Jul,Campbell2024Mar,BibEntry2007Jan,Nielsen2010Dec}. The other is quantum error mitigation (QEM): finding ways to reduce or cancel errors in today’s small to intermediate-scale quantum devices without the full overhead of error correction \cite{Cai2023Dec}. 
This article briefly reviews these strategies and highlights the pivotal contributions of Canadian institutions and researchers, from pioneering theoretical codes to cutting-edge experiments and collaborations (see geographical distribution in Fig.~\ref{fig:Canada_heatmap}).

\section*{Early Canadian Breakthroughs in Quantum Error Correction}
\begin{figure*}[t]
        \centering
        \includegraphics[width=1.0\linewidth]{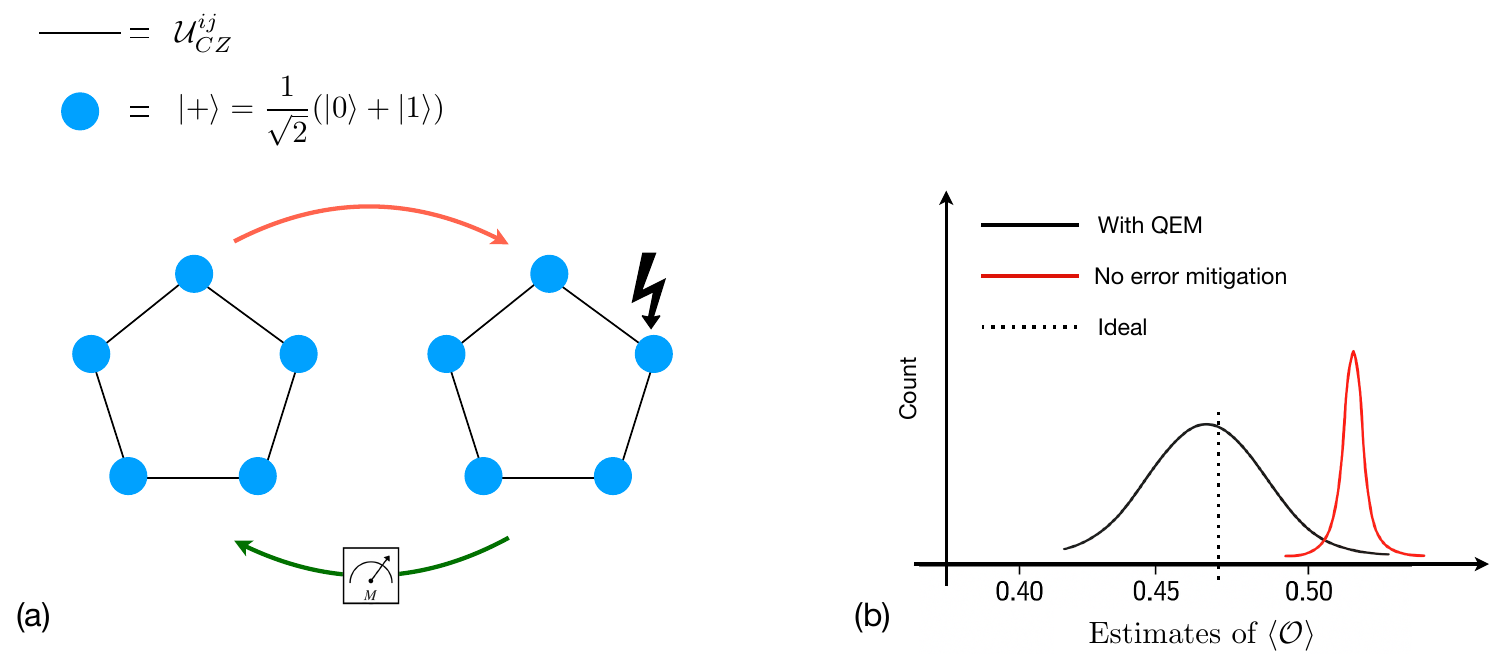}
        \caption{(a) The five-qubit quantum error correcting code, the smallest QECC that can correct one arbitrary quantum error, can be constructed by first creating all five qubits in $\ket{+}$ state and apply $\mathcal{U}_{CZ}$ gate in between each qubit.
        The resulting pentagon can now be considered as one logical qubit. Whenever there is an error in one of the five physical qubits (as indicated by the red arrow), one can detect via syndrome measurements and correct it back as indicated by the green arrow. Readers are referred to Ref. \cite{Kyaw2015Feb} for some realistic schemes to create such a QECC with superconducting qubits. 
        (b) Distributions of estimating some quantum observable $\mathcal{O}$ without any quantum error mitigation scheme (red curve) and with quantum error mitigation (black curve). Ideal expected outcome is indicated with a dotted vertical line. 
    }
        \label{fig:main_fig}
\end{figure*}

The theoretical roots of QEC run through Quebec. In 1996, Laflamme et al. unveiled the five-qubit “perfect” code, the smallest possible scheme that corrects any single-qubit error \cite{Laflamme1996Jul,Bennett1996Nov} (see \figref{fig:main_fig}(a)). Their work produced the Knill–Laflamme conditions \cite{Knill1996Apr}, standard criteria for deciding whether a set of states forms a valid quantum code. Two years later, Laflamme, NMR pioneer David Cory and collaborators delivered the first laboratory proof-of-principle demonstration of QEC, encoding information into the nuclear spins of a liquid molecule and successfully reversing an induced error \cite{Cory1998Sep}.
Daniel Gottesman, who was affiliated with the Perimeter Institute, formalized stabilizer codes \cite{Gottesman1997} and co-invented the Gottesman–Kitaev–Preskill (GKP) code for continuous-variable quantum systems \cite{Gottesman2001Jun}. 
This was one of the significant developments in this field and is now the main driving force behind many of the qumode quantum information processing~\cite{Konno2024Jan}.
David Poulin in Sherbrooke developed efficient decoding algorithms and showed, via threshold theorems, that if each physical qubit’s error drops below $\sim 1$\%, scaling up will actually make logical errors rarer \cite{Poulin2006Nov}.
With the development of the theory of quantum error correction codes and the stabilizer formalism, Canadian and Canada-affiliated researchers played a pivotal role in advancing this field during the first decade of the 21st century~\cite{PhysRevLett.85.1758, doi:10.1142/S0219749904000079, PhysRevLett.95.180501, PhysRevLett.94.180501, Poulin2006Nov, 10.5555/2011665.2011666, PhysRevLett.98.190504, Raussendorf_2007,  Broadbent_UBQC, PhysRevA.80.052312,
DBLP:journals/tit/PoulinTO09, PhysRevLett.104.050504,9259940, Koenig2010Dec}.

This interplay between elegant mathematics and hands-on experiment set the stage for global efforts to build surface-code patches; two-dimensional grids whose local checks catch error “ripples.” Google’s 2023 experiment used 49 superconducting qubits to form a distance-5 surface code whose logical error rate beat that of a smaller code, validating decades of theory \cite{BibEntry2023Feb}.

\section*{Bridging Qubits and Codes: Canadian Advances in Quantum Error Correction and Detection}

As a result of early foundational achievements, Canada has rapidly become a hotbed for innovation in QEC. This includes industry, academic, and government labs that are tackling the challenge from both the code design and hardware implementation fronts. A unifying theme in these efforts is a drive to reduce the daunting resource overhead required by traditional QEC schemes (like the surface code) while maintaining or improving error suppression capabilities. These Canadian contributions range from novel QEC codes and software tools to experimental breakthroughs towards the realization of fault-tolerant hardware, all in service of pushing quantum computers toward practical, error-resilient operation. 

Below we give our handpicked highlights of these innovations, which are by no means exhaustive:

\begin{enumerate}
    \item \textit{Towards Fault-tolerant QC}: Xanadu has focused on bosonic encoding strategies, notably using GKP photonic qubits, to simplify quantum error correction requirements. In recent work, they demonstrated how universal quantum error correction could be implemented with significantly reduced qubit overhead using linear optical operations~\cite{Walshe2025}. Supported by government investment, Xanadu is actively working toward a fully fault-tolerant photonic quantum computing platform, highlighting the strong integration of theoretical and hardware efforts within Canada~\cite{Trudeau2023}.

    \item \textit{Advancements in quantum low-density parity-check codes}: One notable recent advancement from the startup Photonic Inc. in Vancouver, which introduced a new family of quantum low-density parity-check (QLDPC) codes called Subsystem Hypergraph Product Simplex codes (SHYPS)~\cite{Malcolm2025}. These codes can significantly reduce the number of physical qubits required for fault-tolerant quantum computing compared to conventional codes such as the surface code. Photonic Inc. further complemented their theoretical work with a specialized hardware architecture optimized for these highly connected QLDPC codes, leveraging photonic technology for efficient entanglement distribution~\cite{Malcolm2025}. Such codes are optimized for use with Photonic Inc.'s hybrid photonic spin-qubit chips.
    In fact, local and global connectivity of physical qubits define the resulting quantum codes \cite{Baspin2022May}.

    \item \textit{Bosonic codes on superconducting hardware}: In superconducting circuits, Sherbrooke-based startup Nord Quantique has made remarkable strides by experimentally demonstrating a hardware-efficient quantum error correction approach using (beyond GKP) bosonic grid states. Their pioneering experiments have shown that this method can actively extend qubit coherence times without excessive redundancy, an essential step towards scalable quantum computing~\cite{Lemonde2024, Lachance2023Oct,Puri2020Aug}. These developments were strongly supported by collaborations within the Sherbrooke quantum ecosystem, including the Institut Quantique and IBM Quantum Hub.

    \item \textit{Surface codes}: Researchers at the University of Waterloo’s Institute for Quantum Computing (IQC) and the Perimeter Institute have introduced groundbreaking QEC codes like the three-dimensional subsystem toric code. This code notably allows single-shot quantum error correction, significantly simplifying the correction process and reducing the overhead needed for practical fault-tolerant quantum computing~\cite{Kubica2022}.
    An experimental realization of distance-three surface code using $17$ physical qubits can also be found in \cite{Krinner2022May}.
    
    \item \textit{Error detection for early-term fault tolerance}: Researchers from the BEIT group have proposed a resource-efficient framework for quantum error detection (QED) tailored to near-term quantum devices~\cite{ginsberg2025quantumerrordetectionearly}. Through simulations of Grover's algorithm under a circuit-level noise model, they demonstrate that optimizing syndrome measurement schedules can enhance algorithm success probabilities. Additionally, they introduce a data-driven method to predict optimal compilation parameters based on circuit and noise characteristics. This work provides actionable guidelines for implementing QED in early-term quantum experiments and underscores its potential as a pragmatic error mitigation strategy for shallow quantum algorithms~\cite{ginsberg2025quantumerrordetectionearly}.

    \item \textit{Neural decoders for topological codes}: Researchers at the University of Waterloo and the Perimeter Institute have pioneered the use of machine learning for quantum error correction by developing a neural network-based decoder. It was demonstrated that a restricted Boltzmann machine could efficiently decode syndrome data from topological quantum codes without requiring explicit algorithmic construction, adapting flexibly to the error distributions encountered. This method has opened a significant new avenue for scalable, adaptive decoding algorithms~\cite{Torlai2017}.

    \item \textit{Flag fault-tolerant error correction}: A new fault-tolerant error correction method applicable to arbitrary-distance quantum codes through the innovative use of flag qubits was introduced. By strategically employing flag qubits to signal problematic syndrome measurements, their approach effectively manages high-weight errors while significantly reducing resource overhead compared to traditional fault-tolerant ancilla techniques ~\cite{Chamberland2018}.

\end{enumerate}

Collectively, these initiatives, ranging from innovative new codes and sophisticated decoding algorithms to cutting-edge hardware implementations, represent a comprehensive and cohesive approach to quantum error correction on Canadian soil. By closely aligning theoretical breakthroughs with practical experimental demonstrations, Canada is playing a crucial role in paving the way for fault-tolerant quantum computing.

\section*{From Codes to Quick Fixes: Canadian Leaders in Quantum Error Mitigation}
Full quantum error correction demands thousands of physical qubits per logical qubit, a resource requirement beyond near-term capabilities. This limitation has led to \textit{quantum error mitigation}, a technique designed to extract greater accuracy from noisy quantum devices with minimal additional resources (see \figref{fig:main_fig}(b)). Canadian research teams have played a key role in advancing several prominent error mitigation methods:

\begin{enumerate}
    \item \textit{Zero-noise extrapolation (ZNE)}: ZNE is a popular method that consists of running the same quantum circuit at different noise levels (using techniques such as stretching gate durations) and mathematically extrapolating to the zero-noise limit. 
    Canadian researchers have been instrumental in developing and improving many error mitigation protocols based on ZNE. Digital ZNE provides a practical framework where unitary folding and parameterized noise scaling can be applied using only gate-level access common to most quantum instruction sets~\cite{9259940}. Another scheme, called variable-noise Clifford data regression (vnCDR), developed with the participation of the University of Waterloo, combines ZNE with Clifford data regression to mitigate errors, was shown to be more powerful than the individual methods of ZNE and CDR~\cite{PhysRevResearch.3.033098}.
    QEM protocols for quantum annealing using ZNE were introduced by a team comprising of researchers from the D-Wave, the University of British Columbia, and Simon-Fraser University~\cite{Raymond2025Mar}. 
    A recent study comprising of researchers from the University of Waterloo and the Vector Institute
    has also shown that ZNE is not the best option for mitigating errors in quantum sensing~\cite{IJAZ2025100042}.
    
    \item \textit{Randomised compiling (RC)}: Conceived by researchers at the University of Waterloo, RC inserts randomly chosen gate twirls that convert coherent, bias-inducing errors into easier-to-model stochastic noise~\cite{PhysRevA.94.052325}. Their spin-off, Quantum Benchmark—later acquired by Keysight—made RC diagnostics widely available~\cite{PhysRevX.11.041039, winick2022conceptsconditionserrorsuppression, beale2023subsystemmeasurements}.
    \item \textit{Symmetry-based post-selection and superposed mitigation}: When a problem conserves a quantity (say, particle number), measurement results that violate this conservation law can be discarded. Building on that idea, a Waterloo research group proposed an approach where different noisy runs interfere quantum-mechanically to cancel certain error terms, a concept now under experimental investigation~\cite{Miguel_Ramiro_2023}.
    \item \textit{Quasi-probabilistic methods}: Error mitigation strategies such as Probabilistic Error Cancellation (PEC) based on quasi-probabilistic methods are among the widely used methods and promise an unbiased result, albeit at the cost of an exponential runtime. Canadian researchers have developed quasi-probabilistic EM techniques that outperform PEC. \textit{Pauli Error Cancellation}~\cite{Ferracin2024efficiently} introduced by researchers from Keysight technologies and the University of Waterloo, was shown to mitigate non-local and gate-dependent noise.
    A constant runtime quasi-probabilistic EM protocol, called \textit{Error Mitigation by Restricted Evolution (EMRE)}, was recently introduced which works by restricting the evolution of the input state~\cite{saxena2024errormitigationrestrictedevolution}. This constant runtime comes at the cost of a small finite bias.
    Another scheme called Hybrid EMRE or HEMRE was also proposed by combining EMRE with PEC. It was shown that using HEMRE, a user can fix the maximum tolerable bias and achieve the mitigated result with a runtime not greater than that of PEC.
    The EMRE framework was subsequently extended to the design of a resource-efficient mitigation protocol based on noise amplification and robust extrapolation to the zero noise limit and is termed as the \textit{Physics-Inspired Extrapolation (PIE)} method~\cite{díezvalle_2025}. Unlike typical ZNE methods, PIE analytically justifies the extrapolation function and assigns operational meaning to the fitting parameters. The method demonstrates good accuracy and robustness in simulating quantum dynamics up to 84 qubits.

    \item \textit{Neural error mitigation}: A team of researchers from 1QBit, the University of Waterloo, the Vector Institute, and the Perimeter Institute for Theoretical Physics introduced a machine learning-based framework for error mitigation in quantum simulations. By leveraging neural networks, the method achieves improved accuracy in quantum chemistry and lattice gauge theory simulations on noisy quantum devices~\cite{Bennewitz_2022}. In another work, researchers introduce a neural model that achieves quantum error mitigation without any prior knowledge of the noise and without training on noise-free data~\cite{Liao2025Jan}.
    \item \textit{QEM on near-term quantum photonic devices}: Photon losses are very common sources of errors in scalable photonic quantum information processing. To address this problem, researchers from Xanadu have developed schemes to mitigate the effects of photon losses in Gaussian boson sampling devices~\cite{Su2021errormitigationnear}.
    
    \item \textit{Other EM strategies}: Other error mitigation schemes and analysis have been introduced by a team of researchers comprising of researchers from Canada. A benchmarking study to understand the application of quantum error mitigation in quantum chemistry was studied in~\cite{D3CP03523A}.
    Another work~\cite{hagge2023errormitigationerrordetection} introduced an error detection scheme by using the Bravyi-Kitaev superfast encoding and showed that it can be used to mitigate errors in quantum chemistry simulations.
    Researchers from Canadian institutions have also been involved in correctly characterizing state preparation and measurement (SPAM) errors which helped both in mitigating such errors and in designing reliable quantum processing units~\cite{PhysRevResearch.3.033285}.
    Lastly, \textit{Mitiq}---a Python package, widely used in the quantum community  to implement and deploy error mitigation protocols---was also developed in collaboration with Canadian researchers~\cite{LaRose2022mitiqsoftware}.
\end{enumerate}
Thanks to these techniques, small quantum processors have already produced chemically and physically meaningful results, despite raw error rates that would otherwise swamp useful signals.

\section*{Remarks and Well-wishes}

Since the late 1990s, quantum error correction (QEC) has been a foundational pillar in the pursuit of building reliable quantum computers. While significant strides continue toward the ultimate goal of fault-tolerant quantum computing involving millions of physical qubits, quantum error mitigation (QEM) has emerged as an invaluable complementary tool, enabling researchers to obtain trustworthy results from today's noisy quantum hardware.

In this brief review, we have highlighted notable contributions from Canadian and Canada-affiliated researchers in academia and industry in both quantum error correction and mitigation. The selected references presented here represent only a curated subset of Canada's extensive contributions to the field (see \figref{fig:Canada_heatmap}).

Looking ahead, Canadian researchers are poised to play a pivotal role in achieving two key milestones. The first is the long-term goal of experimentally demonstrating quantum error correction beyond the fault-tolerance threshold \textit{at useful scales}. While beyond-threshold computation has now been performed at small scales \cite{BibEntry2023Feb}, we estimate that useful simulations in chemistry, for example, will require thousands of logical qubits to demonstrate practical usefulness~\cite{PRXQuantum.2.030305}. The second is a more immediate milestone in the noisy intermediate-scale quantum (NISQ) and early fault-tolerant quantum computing (FTQC) era, involving the integration of quantum error mitigation techniques with error-correcting codes on existing hardware to enable utility-scale quantum computations~\cite{PRXQuantum.3.010345}.

We find ourselves in a remarkable era, achieving a level of quantum system control and manipulation unimaginable to the pioneers of quantum mechanics, including the ability to remotely operate quantum systems from across the globe. While substantial work remains to elevate noisy quantum devices to practical, industry-relevant applications, the significant progress already made provides ample reason for optimism. As we celebrate the 2025 International Year of Quantum Science and Technology (IYQ), it is fitting to reflect on these achievements and recognize the promising journey ahead.

\section*{Acknowledgements}
We would like to extend our thanks to Yipeng Ji, Paria Nejat of LG Electronics Toronto AI Lab and Sean Kim of LG Electronics, AI Lab, for their constant administrative support. 

\section*{Data Availability}
The data used to generate the Canada heatmap is available upon request.

\bibliography{ref}

\begin{thebibliography}{62}%
\makeatletter
\providecommand \@ifxundefined [1]{%
 \@ifx{#1\undefined}
}%
\providecommand \@ifnum [1]{%
 \ifnum #1\expandafter \@firstoftwo
 \else \expandafter \@secondoftwo
 \fi
}%
\providecommand \@ifx [1]{%
 \ifx #1\expandafter \@firstoftwo
 \else \expandafter \@secondoftwo
 \fi
}%
\providecommand \natexlab [1]{#1}%
\providecommand \enquote  [1]{``#1''}%
\providecommand \bibnamefont  [1]{#1}%
\providecommand \bibfnamefont [1]{#1}%
\providecommand \citenamefont [1]{#1}%
\providecommand \href@noop [0]{\@secondoftwo}%
\providecommand \href [0]{\begingroup \@sanitize@url \@href}%
\providecommand \@href[1]{\@@startlink{#1}\@@href}%
\providecommand \@@href[1]{\endgroup#1\@@endlink}%
\providecommand \@sanitize@url [0]{\catcode `\\12\catcode `\$12\catcode `\&12\catcode `\#12\catcode `\^12\catcode `\_12\catcode `\%12\relax}%
\providecommand \@@startlink[1]{}%
\providecommand \@@endlink[0]{}%
\providecommand \url  [0]{\begingroup\@sanitize@url \@url }%
\providecommand \@url [1]{\endgroup\@href {#1}{\urlprefix }}%
\providecommand \urlprefix  [0]{URL }%
\providecommand \Eprint [0]{\href }%
\providecommand \doibase [0]{https://doi.org/}%
\providecommand \selectlanguage [0]{\@gobble}%
\providecommand \bibinfo  [0]{\@secondoftwo}%
\providecommand \bibfield  [0]{\@secondoftwo}%
\providecommand \translation [1]{[#1]}%
\providecommand \BibitemOpen [0]{}%
\providecommand \bibitemStop [0]{}%
\providecommand \bibitemNoStop [0]{.\EOS\space}%
\providecommand \EOS [0]{\spacefactor3000\relax}%
\providecommand \BibitemShut  [1]{\csname bibitem#1\endcsname}%
\let\auto@bib@innerbib\@empty
\bibitem [{\citenamefont {Martinis}(2021)}]{Martinis2021Jun}%
  \BibitemOpen
  \bibfield  {author} {\bibinfo {author} {\bibfnamefont {J.~M.}\ \bibnamefont {Martinis}},\ }\bibfield  {title} {\bibinfo {title} {{Saving superconducting quantum processors from decay and correlated errors generated by gamma and cosmic rays}},\ }\href {https://doi.org/10.1038/s41534-021-00431-0} {\bibfield  {journal} {\bibinfo  {journal} {npj Quantum Inf.}\ }\textbf {\bibinfo {volume} {7}},\ \bibinfo {pages} {1} (\bibinfo {year} {2021})}\BibitemShut {NoStop}%
\bibitem [{\citenamefont {GoogleQuantumAI}(2023)}]{BibEntry2023Feb}%
  \BibitemOpen
  \bibfield  {author} {\bibinfo {author} {\bibnamefont {GoogleQuantumAI}},\ }\bibfield  {title} {\bibinfo {title} {{Suppressing quantum errors by scaling a surface code logical qubit}},\ }\href {https://doi.org/10.1038/s41586-022-05434-1} {\bibfield  {journal} {\bibinfo  {journal} {Nature}\ }\textbf {\bibinfo {volume} {614}},\ \bibinfo {pages} {676} (\bibinfo {year} {2023})}\BibitemShut {NoStop}%
\bibitem [{\citenamefont {Brun}(2013)}]{BibEntry2013Sep}%
  \BibitemOpen
  \bibfield  {author} {\bibinfo {author} {\bibfnamefont {T.~A.}\ \bibnamefont {Brun}},\ }\bibfield  {title} {\bibinfo {title} {{Quantum Error Correction}},\ }\bibfield  {journal} {\bibinfo  {journal} {Cambridge University Press}\ }\href {https://doi.org/10.1017/CBO9781139034807} {10.1017/CBO9781139034807} (\bibinfo {year} {2013}),\ \bibinfo {note} {[Online; accessed 20. May 2025]}\BibitemShut {NoStop}%
\bibitem [{\citenamefont {Roffe}(2019)}]{Roffe2019Jul}%
  \BibitemOpen
  \bibfield  {author} {\bibinfo {author} {\bibfnamefont {J.}~\bibnamefont {Roffe}},\ }\bibfield  {title} {\bibinfo {title} {{Quantum error correction: an introductory guide}},\ }\href {https://www.tandfonline.com/doi/full/10.1080/00107514.2019.1667078} {\bibfield  {journal} {\bibinfo  {journal} {Contemp. Phys.}\ } (\bibinfo {year} {2019})}\BibitemShut {NoStop}%
\bibitem [{\citenamefont {Campbell}(2024)}]{Campbell2024Mar}%
  \BibitemOpen
  \bibfield  {author} {\bibinfo {author} {\bibfnamefont {E.}~\bibnamefont {Campbell}},\ }\bibfield  {title} {\bibinfo {title} {{A series of fast-paced advances in Quantum Error Correction}},\ }\href {https://doi.org/10.1038/s42254-024-00706-3} {\bibfield  {journal} {\bibinfo  {journal} {Nat. Rev. Phys.}\ }\textbf {\bibinfo {volume} {6}},\ \bibinfo {pages} {160} (\bibinfo {year} {2024})}\BibitemShut {NoStop}%
\bibitem [{\citenamefont {Kaye}\ \emph {et~al.}(2007)\citenamefont {Kaye}, \citenamefont {Laflamme},\ and\ \citenamefont {Mosca}}]{BibEntry2007Jan}%
  \BibitemOpen
  \bibfield  {author} {\bibinfo {author} {\bibfnamefont {P.}~\bibnamefont {Kaye}}, \bibinfo {author} {\bibfnamefont {R.}~\bibnamefont {Laflamme}},\ and\ \bibinfo {author} {\bibfnamefont {M.}~\bibnamefont {Mosca}},\ }\bibfield  {title} {\bibinfo {title} {{An Introduction to Quantum Computing}},\ }\href {https://global.oup.com/academic/product/an-introduction-to-quantum-computing-9780198570493} {\bibfield  {journal} {\bibinfo  {journal} {Oxford University Press}\ } (\bibinfo {year} {2007})}\BibitemShut {NoStop}%
\bibitem [{\citenamefont {Nielsen}\ and\ \citenamefont {Chuang}(2010)}]{Nielsen2010Dec}%
  \BibitemOpen
  \bibfield  {author} {\bibinfo {author} {\bibfnamefont {M.~A.}\ \bibnamefont {Nielsen}}\ and\ \bibinfo {author} {\bibfnamefont {I.~L.}\ \bibnamefont {Chuang}},\ }\bibfield  {title} {\bibinfo {title} {{Quantum Computation and Quantum Information: 10th Anniversary Edition}},\ }\bibfield  {journal} {\bibinfo  {journal} {Cambridge University Press}\ }\href {https://doi.org/10.1017/CBO9780511976667} {10.1017/CBO9780511976667} (\bibinfo {year} {2010})\BibitemShut {NoStop}%
\bibitem [{\citenamefont {Cai}\ \emph {et~al.}(2023)\citenamefont {Cai}, \citenamefont {Babbush}, \citenamefont {Benjamin}, \citenamefont {Endo}, \citenamefont {Huggins}, \citenamefont {Li}, \citenamefont {McClean},\ and\ \citenamefont {O{'}Brien}}]{Cai2023Dec}%
  \BibitemOpen
  \bibfield  {author} {\bibinfo {author} {\bibfnamefont {Z.}~\bibnamefont {Cai}}, \bibinfo {author} {\bibfnamefont {R.}~\bibnamefont {Babbush}}, \bibinfo {author} {\bibfnamefont {S.~C.}\ \bibnamefont {Benjamin}}, \bibinfo {author} {\bibfnamefont {S.}~\bibnamefont {Endo}}, \bibinfo {author} {\bibfnamefont {W.~J.}\ \bibnamefont {Huggins}}, \bibinfo {author} {\bibfnamefont {Y.}~\bibnamefont {Li}}, \bibinfo {author} {\bibfnamefont {J.~R.}\ \bibnamefont {McClean}},\ and\ \bibinfo {author} {\bibfnamefont {T.~E.}\ \bibnamefont {O{'}Brien}},\ }\bibfield  {title} {\bibinfo {title} {{Quantum error mitigation}},\ }\href {https://doi.org/10.1103/RevModPhys.95.045005} {\bibfield  {journal} {\bibinfo  {journal} {Rev. Mod. Phys.}\ }\textbf {\bibinfo {volume} {95}},\ \bibinfo {pages} {045005} (\bibinfo {year} {2023})}\BibitemShut {NoStop}%
\bibitem [{\citenamefont {Laflamme}\ \emph {et~al.}(1996)\citenamefont {Laflamme}, \citenamefont {Miquel}, \citenamefont {Paz},\ and\ \citenamefont {Zurek}}]{Laflamme1996Jul}%
  \BibitemOpen
  \bibfield  {author} {\bibinfo {author} {\bibfnamefont {R.}~\bibnamefont {Laflamme}}, \bibinfo {author} {\bibfnamefont {C.}~\bibnamefont {Miquel}}, \bibinfo {author} {\bibfnamefont {J.~P.}\ \bibnamefont {Paz}},\ and\ \bibinfo {author} {\bibfnamefont {W.~H.}\ \bibnamefont {Zurek}},\ }\bibfield  {title} {\bibinfo {title} {{Perfect Quantum Error Correcting Code}},\ }\href {https://doi.org/10.1103/PhysRevLett.77.198} {\bibfield  {journal} {\bibinfo  {journal} {Phys. Rev. Lett.}\ }\textbf {\bibinfo {volume} {77}},\ \bibinfo {pages} {198} (\bibinfo {year} {1996})}\BibitemShut {NoStop}%
\bibitem [{\citenamefont {Bennett}\ \emph {et~al.}(1996)\citenamefont {Bennett}, \citenamefont {DiVincenzo}, \citenamefont {Smolin},\ and\ \citenamefont {Wootters}}]{Bennett1996Nov}%
  \BibitemOpen
  \bibfield  {author} {\bibinfo {author} {\bibfnamefont {C.~H.}\ \bibnamefont {Bennett}}, \bibinfo {author} {\bibfnamefont {D.~P.}\ \bibnamefont {DiVincenzo}}, \bibinfo {author} {\bibfnamefont {J.~A.}\ \bibnamefont {Smolin}},\ and\ \bibinfo {author} {\bibfnamefont {W.~K.}\ \bibnamefont {Wootters}},\ }\bibfield  {title} {\bibinfo {title} {{Mixed-state entanglement and quantum error correction}},\ }\href {https://doi.org/10.1103/PhysRevA.54.3824} {\bibfield  {journal} {\bibinfo  {journal} {Phys. Rev. A}\ }\textbf {\bibinfo {volume} {54}},\ \bibinfo {pages} {3824} (\bibinfo {year} {1996})}\BibitemShut {NoStop}%
\bibitem [{\citenamefont {Knill}\ and\ \citenamefont {Laflamme}(1996)}]{Knill1996Apr}%
  \BibitemOpen
  \bibfield  {author} {\bibinfo {author} {\bibfnamefont {E.}~\bibnamefont {Knill}}\ and\ \bibinfo {author} {\bibfnamefont {R.}~\bibnamefont {Laflamme}},\ }\bibfield  {title} {\bibinfo {title} {{A Theory of Quantum Error-Correcting Codes}},\ }\bibfield  {journal} {\bibinfo  {journal} {arXiv}\ }\href {https://doi.org/10.1103/PhysRevLett.84.2525} {10.1103/PhysRevLett.84.2525} (\bibinfo {year} {1996}),\ \Eprint {https://arxiv.org/abs/quant-ph/9604034} {quant-ph/9604034} \BibitemShut {NoStop}%
\bibitem [{\citenamefont {Cory}\ \emph {et~al.}(1998)\citenamefont {Cory}, \citenamefont {Price}, \citenamefont {Maas}, \citenamefont {Knill}, \citenamefont {Laflamme}, \citenamefont {Zurek}, \citenamefont {Havel},\ and\ \citenamefont {Somaroo}}]{Cory1998Sep}%
  \BibitemOpen
  \bibfield  {author} {\bibinfo {author} {\bibfnamefont {D.~G.}\ \bibnamefont {Cory}}, \bibinfo {author} {\bibfnamefont {M.~D.}\ \bibnamefont {Price}}, \bibinfo {author} {\bibfnamefont {W.}~\bibnamefont {Maas}}, \bibinfo {author} {\bibfnamefont {E.}~\bibnamefont {Knill}}, \bibinfo {author} {\bibfnamefont {R.}~\bibnamefont {Laflamme}}, \bibinfo {author} {\bibfnamefont {W.~H.}\ \bibnamefont {Zurek}}, \bibinfo {author} {\bibfnamefont {T.~F.}\ \bibnamefont {Havel}},\ and\ \bibinfo {author} {\bibfnamefont {S.~S.}\ \bibnamefont {Somaroo}},\ }\bibfield  {title} {\bibinfo {title} {{Experimental Quantum Error Correction}},\ }\href {https://doi.org/10.1103/PhysRevLett.81.2152} {\bibfield  {journal} {\bibinfo  {journal} {Phys. Rev. Lett.}\ }\textbf {\bibinfo {volume} {81}},\ \bibinfo {pages} {2152} (\bibinfo {year} {1998})}\BibitemShut {NoStop}%
\bibitem [{\citenamefont {Gottesman}(1997)}]{Gottesman1997}%
  \BibitemOpen
  \bibfield  {author} {\bibinfo {author} {\bibfnamefont {D.~E.}\ \bibnamefont {Gottesman}},\ }\bibfield  {title} {\bibinfo {title} {Stabilizer codes and quantum error correction},\ }\href {https://thesis.library.caltech.edu/2900} {\bibfield  {journal} {\bibinfo  {journal} {Caltech PhD thesis}\ } (\bibinfo {year} {1997})}\BibitemShut {NoStop}%
\bibitem [{\citenamefont {Gottesman}\ \emph {et~al.}(2001)\citenamefont {Gottesman}, \citenamefont {Kitaev},\ and\ \citenamefont {Preskill}}]{Gottesman2001Jun}%
  \BibitemOpen
  \bibfield  {author} {\bibinfo {author} {\bibfnamefont {D.}~\bibnamefont {Gottesman}}, \bibinfo {author} {\bibfnamefont {A.}~\bibnamefont {Kitaev}},\ and\ \bibinfo {author} {\bibfnamefont {J.}~\bibnamefont {Preskill}},\ }\bibfield  {title} {\bibinfo {title} {{Encoding a qubit in an oscillator}},\ }\href {https://doi.org/10.1103/PhysRevA.64.012310} {\bibfield  {journal} {\bibinfo  {journal} {Phys. Rev. A}\ }\textbf {\bibinfo {volume} {64}},\ \bibinfo {pages} {012310} (\bibinfo {year} {2001})}\BibitemShut {NoStop}%
\bibitem [{\citenamefont {Konno}\ \emph {et~al.}(2024)\citenamefont {Konno}, \citenamefont {Asavanant}, \citenamefont {Hanamura}, \citenamefont {Nagayoshi}, \citenamefont {Fukui}, \citenamefont {Sakaguchi}, \citenamefont {Ide}, \citenamefont {China}, \citenamefont {Yabuno}, \citenamefont {Miki}, \citenamefont {Terai}, \citenamefont {Takase}, \citenamefont {Endo}, \citenamefont {Marek}, \citenamefont {Filip}, \citenamefont {van Loock},\ and\ \citenamefont {Furusawa}}]{Konno2024Jan}%
  \BibitemOpen
  \bibfield  {author} {\bibinfo {author} {\bibfnamefont {S.}~\bibnamefont {Konno}}, \bibinfo {author} {\bibfnamefont {W.}~\bibnamefont {Asavanant}}, \bibinfo {author} {\bibfnamefont {F.}~\bibnamefont {Hanamura}}, \bibinfo {author} {\bibfnamefont {H.}~\bibnamefont {Nagayoshi}}, \bibinfo {author} {\bibfnamefont {K.}~\bibnamefont {Fukui}}, \bibinfo {author} {\bibfnamefont {A.}~\bibnamefont {Sakaguchi}}, \bibinfo {author} {\bibfnamefont {R.}~\bibnamefont {Ide}}, \bibinfo {author} {\bibfnamefont {F.}~\bibnamefont {China}}, \bibinfo {author} {\bibfnamefont {M.}~\bibnamefont {Yabuno}}, \bibinfo {author} {\bibfnamefont {S.}~\bibnamefont {Miki}}, \bibinfo {author} {\bibfnamefont {H.}~\bibnamefont {Terai}}, \bibinfo {author} {\bibfnamefont {K.}~\bibnamefont {Takase}}, \bibinfo {author} {\bibfnamefont {M.}~\bibnamefont {Endo}}, \bibinfo {author} {\bibfnamefont {P.}~\bibnamefont {Marek}}, \bibinfo {author} {\bibfnamefont {R.}~\bibnamefont {Filip}}, \bibinfo {author} {\bibfnamefont {P.}~\bibnamefont {van Loock}},\ and\
  \bibinfo {author} {\bibfnamefont {A.}~\bibnamefont {Furusawa}},\ }\bibfield  {title} {\bibinfo {title} {{Logical states for fault-tolerant quantum computation with propagating light}},\ }\href {https://doi.org/10.1126/science.adk7560} {\bibfield  {journal} {\bibinfo  {journal} {Science}\ }\textbf {\bibinfo {volume} {383}},\ \bibinfo {pages} {289} (\bibinfo {year} {2024})}\BibitemShut {NoStop}%
\bibitem [{\citenamefont {Poulin}(2006)}]{Poulin2006Nov}%
  \BibitemOpen
  \bibfield  {author} {\bibinfo {author} {\bibfnamefont {D.}~\bibnamefont {Poulin}},\ }\bibfield  {title} {\bibinfo {title} {{Optimal and efficient decoding of concatenated quantum block codes}},\ }\href {https://doi.org/10.1103/PhysRevA.74.052333} {\bibfield  {journal} {\bibinfo  {journal} {Phys. Rev. A}\ }\textbf {\bibinfo {volume} {74}},\ \bibinfo {pages} {052333} (\bibinfo {year} {2006})}\BibitemShut {NoStop}%
\bibitem [{\citenamefont {Bacon}\ \emph {et~al.}(2000)\citenamefont {Bacon}, \citenamefont {Kempe}, \citenamefont {Lidar},\ and\ \citenamefont {Whaley}}]{PhysRevLett.85.1758}%
  \BibitemOpen
  \bibfield  {author} {\bibinfo {author} {\bibfnamefont {D.}~\bibnamefont {Bacon}}, \bibinfo {author} {\bibfnamefont {J.}~\bibnamefont {Kempe}}, \bibinfo {author} {\bibfnamefont {D.~A.}\ \bibnamefont {Lidar}},\ and\ \bibinfo {author} {\bibfnamefont {K.~B.}\ \bibnamefont {Whaley}},\ }\bibfield  {title} {\bibinfo {title} {Universal fault-tolerant quantum computation on decoherence-free subspaces},\ }\href {https://doi.org/10.1103/PhysRevLett.85.1758} {\bibfield  {journal} {\bibinfo  {journal} {Phys. Rev. Lett.}\ }\textbf {\bibinfo {volume} {85}},\ \bibinfo {pages} {1758} (\bibinfo {year} {2000})}\BibitemShut {NoStop}%
\bibitem [{\citenamefont {Grassl}\ \emph {et~al.}(2004)\citenamefont {Grassl}, \citenamefont {Beth},\ and\ \citenamefont {R\"{o}tteler}}]{doi:10.1142/S0219749904000079}%
  \BibitemOpen
  \bibfield  {author} {\bibinfo {author} {\bibfnamefont {M.}~\bibnamefont {Grassl}}, \bibinfo {author} {\bibfnamefont {T.}~\bibnamefont {Beth}},\ and\ \bibinfo {author} {\bibfnamefont {M.}~\bibnamefont {R\"{o}tteler}},\ }\bibfield  {title} {\bibinfo {title} {On optimal quantum codes},\ }\href {https://doi.org/10.1142/S0219749904000079} {\bibfield  {journal} {\bibinfo  {journal} {International Journal of Quantum Information}\ }\textbf {\bibinfo {volume} {02}},\ \bibinfo {pages} {55} (\bibinfo {year} {2004})},\ \Eprint {https://arxiv.org/abs/https://doi.org/10.1142/S0219749904000079} {https://doi.org/10.1142/S0219749904000079} \BibitemShut {NoStop}%
\bibitem [{\citenamefont {Khodjasteh}\ and\ \citenamefont {Lidar}(2005)}]{PhysRevLett.95.180501}%
  \BibitemOpen
  \bibfield  {author} {\bibinfo {author} {\bibfnamefont {K.}~\bibnamefont {Khodjasteh}}\ and\ \bibinfo {author} {\bibfnamefont {D.~A.}\ \bibnamefont {Lidar}},\ }\bibfield  {title} {\bibinfo {title} {Fault-tolerant quantum dynamical decoupling},\ }\href {https://doi.org/10.1103/PhysRevLett.95.180501} {\bibfield  {journal} {\bibinfo  {journal} {Phys. Rev. Lett.}\ }\textbf {\bibinfo {volume} {95}},\ \bibinfo {pages} {180501} (\bibinfo {year} {2005})}\BibitemShut {NoStop}%
\bibitem [{\citenamefont {Kribs}\ \emph {et~al.}(2005)\citenamefont {Kribs}, \citenamefont {Laflamme},\ and\ \citenamefont {Poulin}}]{PhysRevLett.94.180501}%
  \BibitemOpen
  \bibfield  {author} {\bibinfo {author} {\bibfnamefont {D.}~\bibnamefont {Kribs}}, \bibinfo {author} {\bibfnamefont {R.}~\bibnamefont {Laflamme}},\ and\ \bibinfo {author} {\bibfnamefont {D.}~\bibnamefont {Poulin}},\ }\bibfield  {title} {\bibinfo {title} {Unified and generalized approach to quantum error correction},\ }\href {https://doi.org/10.1103/PhysRevLett.94.180501} {\bibfield  {journal} {\bibinfo  {journal} {Phys. Rev. Lett.}\ }\textbf {\bibinfo {volume} {94}},\ \bibinfo {pages} {180501} (\bibinfo {year} {2005})}\BibitemShut {NoStop}%
\bibitem [{\citenamefont {Aliferis}\ \emph {et~al.}(2006)\citenamefont {Aliferis}, \citenamefont {Gottesman},\ and\ \citenamefont {Preskill}}]{10.5555/2011665.2011666}%
  \BibitemOpen
  \bibfield  {author} {\bibinfo {author} {\bibfnamefont {P.}~\bibnamefont {Aliferis}}, \bibinfo {author} {\bibfnamefont {D.}~\bibnamefont {Gottesman}},\ and\ \bibinfo {author} {\bibfnamefont {J.}~\bibnamefont {Preskill}},\ }\bibfield  {title} {\bibinfo {title} {Quantum accuracy threshold for concatenated distance-3 codes},\ }\href@noop {} {\bibfield  {journal} {\bibinfo  {journal} {Quantum Info. Comput.}\ }\textbf {\bibinfo {volume} {6}},\ \bibinfo {pages} {97–165} (\bibinfo {year} {2006})}\BibitemShut {NoStop}%
\bibitem [{\citenamefont {Raussendorf}\ and\ \citenamefont {Harrington}(2007)}]{PhysRevLett.98.190504}%
  \BibitemOpen
  \bibfield  {author} {\bibinfo {author} {\bibfnamefont {R.}~\bibnamefont {Raussendorf}}\ and\ \bibinfo {author} {\bibfnamefont {J.}~\bibnamefont {Harrington}},\ }\bibfield  {title} {\bibinfo {title} {Fault-tolerant quantum computation with high threshold in two dimensions},\ }\href {https://doi.org/10.1103/PhysRevLett.98.190504} {\bibfield  {journal} {\bibinfo  {journal} {Phys. Rev. Lett.}\ }\textbf {\bibinfo {volume} {98}},\ \bibinfo {pages} {190504} (\bibinfo {year} {2007})}\BibitemShut {NoStop}%
\bibitem [{\citenamefont {Raussendorf}\ \emph {et~al.}(2007)\citenamefont {Raussendorf}, \citenamefont {Harrington},\ and\ \citenamefont {Goyal}}]{Raussendorf_2007}%
  \BibitemOpen
  \bibfield  {author} {\bibinfo {author} {\bibfnamefont {R.}~\bibnamefont {Raussendorf}}, \bibinfo {author} {\bibfnamefont {J.}~\bibnamefont {Harrington}},\ and\ \bibinfo {author} {\bibfnamefont {K.}~\bibnamefont {Goyal}},\ }\bibfield  {title} {\bibinfo {title} {Topological fault-tolerance in cluster state quantum computation},\ }\href {https://doi.org/10.1088/1367-2630/9/6/199} {\bibfield  {journal} {\bibinfo  {journal} {New Journal of Physics}\ }\textbf {\bibinfo {volume} {9}},\ \bibinfo {pages} {199} (\bibinfo {year} {2007})}\BibitemShut {NoStop}%
\bibitem [{\citenamefont {Broadbent}\ \emph {et~al.}(2009)\citenamefont {Broadbent}, \citenamefont {Fitzsimons},\ and\ \citenamefont {Kashefi}}]{Broadbent_UBQC}%
  \BibitemOpen
  \bibfield  {author} {\bibinfo {author} {\bibfnamefont {A.}~\bibnamefont {Broadbent}}, \bibinfo {author} {\bibfnamefont {J.}~\bibnamefont {Fitzsimons}},\ and\ \bibinfo {author} {\bibfnamefont {E.}~\bibnamefont {Kashefi}},\ }\bibfield  {title} {\bibinfo {title} {Universal blind quantum computation},\ }\href {https://doi.org/10.1109/FOCS.2009.36} {\bibfield  {journal} {\bibinfo  {journal} {2009 50th Annual IEEE Symposium on Foundations of Computer Science}\ ,\ \bibinfo {pages} {517}} (\bibinfo {year} {2009})}\BibitemShut {NoStop}%
\bibitem [{\citenamefont {Fowler}\ \emph {et~al.}(2009)\citenamefont {Fowler}, \citenamefont {Stephens},\ and\ \citenamefont {Groszkowski}}]{PhysRevA.80.052312}%
  \BibitemOpen
  \bibfield  {author} {\bibinfo {author} {\bibfnamefont {A.~G.}\ \bibnamefont {Fowler}}, \bibinfo {author} {\bibfnamefont {A.~M.}\ \bibnamefont {Stephens}},\ and\ \bibinfo {author} {\bibfnamefont {P.}~\bibnamefont {Groszkowski}},\ }\bibfield  {title} {\bibinfo {title} {High-threshold universal quantum computation on the surface code},\ }\href {https://doi.org/10.1103/PhysRevA.80.052312} {\bibfield  {journal} {\bibinfo  {journal} {Phys. Rev. A}\ }\textbf {\bibinfo {volume} {80}},\ \bibinfo {pages} {052312} (\bibinfo {year} {2009})}\BibitemShut {NoStop}%
\bibitem [{\citenamefont {Poulin}\ \emph {et~al.}(2009)\citenamefont {Poulin}, \citenamefont {Tillich},\ and\ \citenamefont {Ollivier}}]{DBLP:journals/tit/PoulinTO09}%
  \BibitemOpen
  \bibfield  {author} {\bibinfo {author} {\bibfnamefont {D.}~\bibnamefont {Poulin}}, \bibinfo {author} {\bibfnamefont {J.}~\bibnamefont {Tillich}},\ and\ \bibinfo {author} {\bibfnamefont {H.}~\bibnamefont {Ollivier}},\ }\bibfield  {title} {\bibinfo {title} {Quantum serial turbo codes},\ }\href {https://doi.org/10.1109/TIT.2009.2018339} {\bibfield  {journal} {\bibinfo  {journal} {{IEEE} Trans. Inf. Theory}\ }\textbf {\bibinfo {volume} {55}},\ \bibinfo {pages} {2776} (\bibinfo {year} {2009})}\BibitemShut {NoStop}%
\bibitem [{\citenamefont {Duclos-Cianci}\ and\ \citenamefont {Poulin}(2010)}]{PhysRevLett.104.050504}%
  \BibitemOpen
  \bibfield  {author} {\bibinfo {author} {\bibfnamefont {G.}~\bibnamefont {Duclos-Cianci}}\ and\ \bibinfo {author} {\bibfnamefont {D.}~\bibnamefont {Poulin}},\ }\bibfield  {title} {\bibinfo {title} {Fast decoders for topological quantum codes},\ }\href {https://doi.org/10.1103/PhysRevLett.104.050504} {\bibfield  {journal} {\bibinfo  {journal} {Phys. Rev. Lett.}\ }\textbf {\bibinfo {volume} {104}},\ \bibinfo {pages} {050504} (\bibinfo {year} {2010})}\BibitemShut {NoStop}%
\bibitem [{\citenamefont {Giurgica-Tiron}\ \emph {et~al.}(2020)\citenamefont {Giurgica-Tiron}, \citenamefont {Hindy}, \citenamefont {LaRose}, \citenamefont {Mari},\ and\ \citenamefont {Zeng}}]{9259940}%
  \BibitemOpen
  \bibfield  {author} {\bibinfo {author} {\bibfnamefont {T.}~\bibnamefont {Giurgica-Tiron}}, \bibinfo {author} {\bibfnamefont {Y.}~\bibnamefont {Hindy}}, \bibinfo {author} {\bibfnamefont {R.}~\bibnamefont {LaRose}}, \bibinfo {author} {\bibfnamefont {A.}~\bibnamefont {Mari}},\ and\ \bibinfo {author} {\bibfnamefont {W.~J.}\ \bibnamefont {Zeng}},\ }\bibfield  {title} {\bibinfo {title} {Digital zero noise extrapolation for quantum error mitigation},\ }\href {https://doi.org/10.1109/QCE49297.2020.00045} {\bibfield  {journal} {\bibinfo  {journal} {IEEE International Conference on Quantum Computing and Engineering (QCE)}\ ,\ \bibinfo {pages} {306}} (\bibinfo {year} {2020})}\BibitemShut {NoStop}%
\bibitem [{\citenamefont {Koenig}\ \emph {et~al.}(2010)\citenamefont {Koenig}, \citenamefont {Kuperberg},\ and\ \citenamefont {Reichardt}}]{Koenig2010Dec}%
  \BibitemOpen
  \bibfield  {author} {\bibinfo {author} {\bibfnamefont {R.}~\bibnamefont {Koenig}}, \bibinfo {author} {\bibfnamefont {G.}~\bibnamefont {Kuperberg}},\ and\ \bibinfo {author} {\bibfnamefont {B.~W.}\ \bibnamefont {Reichardt}},\ }\bibfield  {title} {\bibinfo {title} {{Quantum computation with Turaev{\textendash}Viro codes}},\ }\href {https://doi.org/10.1016/j.aop.2010.08.001} {\bibfield  {journal} {\bibinfo  {journal} {Ann. Phys.}\ }\textbf {\bibinfo {volume} {325}},\ \bibinfo {pages} {2707} (\bibinfo {year} {2010})}\BibitemShut {NoStop}%
\bibitem [{\citenamefont {Kyaw}\ \emph {et~al.}(2015)\citenamefont {Kyaw}, \citenamefont {Herrera-Mart{\ifmmode\acute{\imath}\else\'{\i}\fi}}, \citenamefont {Solano}, \citenamefont {Romero},\ and\ \citenamefont {Kwek}}]{Kyaw2015Feb}%
  \BibitemOpen
  \bibfield  {author} {\bibinfo {author} {\bibfnamefont {T.~H.}\ \bibnamefont {Kyaw}}, \bibinfo {author} {\bibfnamefont {D.~A.}\ \bibnamefont {Herrera-Mart{\ifmmode\acute{\imath}\else\'{\i}\fi}}}, \bibinfo {author} {\bibfnamefont {E.}~\bibnamefont {Solano}}, \bibinfo {author} {\bibfnamefont {G.}~\bibnamefont {Romero}},\ and\ \bibinfo {author} {\bibfnamefont {L.-C.}\ \bibnamefont {Kwek}},\ }\bibfield  {title} {\bibinfo {title} {{Creation of quantum error correcting codes in the ultrastrong coupling regime}},\ }\href {https://doi.org/10.1103/PhysRevB.91.064503} {\bibfield  {journal} {\bibinfo  {journal} {Phys. Rev. B}\ }\textbf {\bibinfo {volume} {91}},\ \bibinfo {pages} {064503} (\bibinfo {year} {2015})}\BibitemShut {NoStop}%
\bibitem [{\citenamefont {Walshe}\ \emph {et~al.}(2025)\citenamefont {Walshe}, \citenamefont {Baragiola}, \citenamefont {Ferretti}, \citenamefont {Gefaell}, \citenamefont {Vasmer}, \citenamefont {Weil}, \citenamefont {Matsuura}, \citenamefont {Jaeken}, \citenamefont {Pantaleoni}, \citenamefont {Han}, \citenamefont {Hillmann}, \citenamefont {Menicucci}, \citenamefont {Tzitrin},\ and\ \citenamefont {Alexander}}]{Walshe2025}%
  \BibitemOpen
  \bibfield  {author} {\bibinfo {author} {\bibfnamefont {B.~W.}\ \bibnamefont {Walshe}}, \bibinfo {author} {\bibfnamefont {B.~Q.}\ \bibnamefont {Baragiola}}, \bibinfo {author} {\bibfnamefont {H.}~\bibnamefont {Ferretti}}, \bibinfo {author} {\bibfnamefont {J.}~\bibnamefont {Gefaell}}, \bibinfo {author} {\bibfnamefont {M.}~\bibnamefont {Vasmer}}, \bibinfo {author} {\bibfnamefont {R.}~\bibnamefont {Weil}}, \bibinfo {author} {\bibfnamefont {T.}~\bibnamefont {Matsuura}}, \bibinfo {author} {\bibfnamefont {T.}~\bibnamefont {Jaeken}}, \bibinfo {author} {\bibfnamefont {G.}~\bibnamefont {Pantaleoni}}, \bibinfo {author} {\bibfnamefont {Z.}~\bibnamefont {Han}}, \bibinfo {author} {\bibfnamefont {T.}~\bibnamefont {Hillmann}}, \bibinfo {author} {\bibfnamefont {N.~C.}\ \bibnamefont {Menicucci}}, \bibinfo {author} {\bibfnamefont {I.}~\bibnamefont {Tzitrin}},\ and\ \bibinfo {author} {\bibfnamefont {R.~N.}\ \bibnamefont {Alexander}},\ }\bibfield  {title} {\bibinfo {title} {Linear-optical quantum computation with arbitrary
  error-correcting codes},\ }\href {https://doi.org/10.1103/PhysRevLett.134.100602} {\bibfield  {journal} {\bibinfo  {journal} {Physical Review Letters}\ }\textbf {\bibinfo {volume} {134}},\ \bibinfo {pages} {100602} (\bibinfo {year} {2025})},\ \bibinfo {note} {demonstrates photonic GKP-based QEC with reduced qubit overhead}\BibitemShut {NoStop}%
\bibitem [{\citenamefont {{Government of Canada}}(2023)}]{Trudeau2023}%
  \BibitemOpen
  \bibfield  {author} {\bibinfo {author} {\bibnamefont {{Government of Canada}}},\ }\href@noop {} {\bibinfo {title} {{Supporting Canada’s leadership in quantum computing to grow the economy and create jobs}}},\ \bibinfo {howpublished} {\url{https://pm.gc.ca/en/news/news-releases/2023/01/23/supporting-canadas-leadership-quantum-computing-grow-economy-and}} (\bibinfo {year} {2023}),\ \bibinfo {note} {prime Minister’s Office press release announcing funding for Xanadu’s fault-tolerant quantum computer}\BibitemShut {NoStop}%
\bibitem [{\citenamefont {Malcolm}\ \emph {et~al.}(2025)\citenamefont {Malcolm}, \citenamefont {Glaudell}, \citenamefont {Fuentes}, \citenamefont {Chandra}, \citenamefont {Schotte}, \citenamefont {DeLisle}, \citenamefont {Haenel}, \citenamefont {Ebrahimi}, \citenamefont {Roffe}, \citenamefont {Quintavalle}, \citenamefont {Beale}, \citenamefont {Lee-Hone},\ and\ \citenamefont {Simmons}}]{Malcolm2025}%
  \BibitemOpen
  \bibfield  {author} {\bibinfo {author} {\bibfnamefont {A.~J.}\ \bibnamefont {Malcolm}}, \bibinfo {author} {\bibfnamefont {A.~N.}\ \bibnamefont {Glaudell}}, \bibinfo {author} {\bibfnamefont {P.}~\bibnamefont {Fuentes}}, \bibinfo {author} {\bibfnamefont {D.}~\bibnamefont {Chandra}}, \bibinfo {author} {\bibfnamefont {A.}~\bibnamefont {Schotte}}, \bibinfo {author} {\bibfnamefont {C.}~\bibnamefont {DeLisle}}, \bibinfo {author} {\bibfnamefont {R.}~\bibnamefont {Haenel}}, \bibinfo {author} {\bibfnamefont {A.}~\bibnamefont {Ebrahimi}}, \bibinfo {author} {\bibfnamefont {J.}~\bibnamefont {Roffe}}, \bibinfo {author} {\bibfnamefont {A.~O.}\ \bibnamefont {Quintavalle}}, \bibinfo {author} {\bibfnamefont {S.~J.}\ \bibnamefont {Beale}}, \bibinfo {author} {\bibfnamefont {N.~R.}\ \bibnamefont {Lee-Hone}},\ and\ \bibinfo {author} {\bibfnamefont {S.}~\bibnamefont {Simmons}},\ }\bibfield  {title} {\bibinfo {title} {Computing efficiently in qldpc codes},\ }\href@noop {} {\bibfield  {journal} {\bibinfo  {journal} {arXiv preprint
  arXiv:2502.07150}\ } (\bibinfo {year} {2025})},\ \bibinfo {note} {photonic Inc. technical report; new SHYPS QLDPC code family}\BibitemShut {NoStop}%
\bibitem [{\citenamefont {Baspin}\ and\ \citenamefont {Krishna}(2022)}]{Baspin2022May}%
  \BibitemOpen
  \bibfield  {author} {\bibinfo {author} {\bibfnamefont {N.}~\bibnamefont {Baspin}}\ and\ \bibinfo {author} {\bibfnamefont {A.}~\bibnamefont {Krishna}},\ }\bibfield  {title} {\bibinfo {title} {{Connectivity constrains quantum codes}},\ }\href {https://doi.org/10.22331/q-2022-05-13-711} {\bibfield  {journal} {\bibinfo  {journal} {Quantum}\ }\textbf {\bibinfo {volume} {6}},\ \bibinfo {pages} {711} (\bibinfo {year} {2022})},\ \Eprint {https://arxiv.org/abs/2106.00765v4} {2106.00765v4} \BibitemShut {NoStop}%
\bibitem [{\citenamefont {Lemonde}\ \emph {et~al.}(2024)\citenamefont {Lemonde}, \citenamefont {Lachance-Quirion}, \citenamefont {Duclos-Cianci}, \citenamefont {Frattini}, \citenamefont {Hopfmueller}, \citenamefont {Gauvin-Ndiaye}, \citenamefont {Camirand-Lemyre},\ and\ \citenamefont {St-Jean}}]{Lemonde2024}%
  \BibitemOpen
  \bibfield  {author} {\bibinfo {author} {\bibfnamefont {M.-A.}\ \bibnamefont {Lemonde}}, \bibinfo {author} {\bibfnamefont {D.}~\bibnamefont {Lachance-Quirion}}, \bibinfo {author} {\bibfnamefont {G.}~\bibnamefont {Duclos-Cianci}}, \bibinfo {author} {\bibfnamefont {N.~E.}\ \bibnamefont {Frattini}}, \bibinfo {author} {\bibfnamefont {F.}~\bibnamefont {Hopfmueller}}, \bibinfo {author} {\bibfnamefont {C.}~\bibnamefont {Gauvin-Ndiaye}}, \bibinfo {author} {\bibfnamefont {J.}~\bibnamefont {Camirand-Lemyre}},\ and\ \bibinfo {author} {\bibfnamefont {P.}~\bibnamefont {St-Jean}},\ }\bibfield  {title} {\bibinfo {title} {Hardware-efficient fault tolerant quantum computing with bosonic grid states in superconducting circuits},\ }\href@noop {} {\bibfield  {journal} {\bibinfo  {journal} {arXiv preprint arXiv:2409.05813}\ } (\bibinfo {year} {2024})},\ \bibinfo {note} {nord Quantique’s approach using GKP bosonic codes for fault tolerance}\BibitemShut {NoStop}%
\bibitem [{\citenamefont {Lachance-Quirion}\ \emph {et~al.}(2023)\citenamefont {Lachance-Quirion}, \citenamefont {Lemonde}, \citenamefont {Simoneau}, \citenamefont {St-Jean}, \citenamefont {Lemieux}, \citenamefont {Turcotte}, \citenamefont {Wright}, \citenamefont {Lacroix}, \citenamefont {Fr{\ifmmode\acute{e}\else\'{e}\fi}chette-Viens}, \citenamefont {Shillito}, \citenamefont {Hopfmueller}, \citenamefont {Tremblay}, \citenamefont {Frattini}, \citenamefont {Lemyre},\ and\ \citenamefont {St-Jean}}]{Lachance2023Oct}%
  \BibitemOpen
  \bibfield  {author} {\bibinfo {author} {\bibfnamefont {D.}~\bibnamefont {Lachance-Quirion}}, \bibinfo {author} {\bibfnamefont {M.-A.}\ \bibnamefont {Lemonde}}, \bibinfo {author} {\bibfnamefont {J.~O.}\ \bibnamefont {Simoneau}}, \bibinfo {author} {\bibfnamefont {L.}~\bibnamefont {St-Jean}}, \bibinfo {author} {\bibfnamefont {P.}~\bibnamefont {Lemieux}}, \bibinfo {author} {\bibfnamefont {S.}~\bibnamefont {Turcotte}}, \bibinfo {author} {\bibfnamefont {W.}~\bibnamefont {Wright}}, \bibinfo {author} {\bibfnamefont {A.}~\bibnamefont {Lacroix}}, \bibinfo {author} {\bibfnamefont {J.}~\bibnamefont {Fr{\ifmmode\acute{e}\else\'{e}\fi}chette-Viens}}, \bibinfo {author} {\bibfnamefont {R.}~\bibnamefont {Shillito}}, \bibinfo {author} {\bibfnamefont {F.}~\bibnamefont {Hopfmueller}}, \bibinfo {author} {\bibfnamefont {M.}~\bibnamefont {Tremblay}}, \bibinfo {author} {\bibfnamefont {N.~E.}\ \bibnamefont {Frattini}}, \bibinfo {author} {\bibfnamefont {J.~C.}\ \bibnamefont {Lemyre}},\ and\ \bibinfo {author} {\bibfnamefont
  {P.}~\bibnamefont {St-Jean}},\ }\bibfield  {title} {\bibinfo {title} {{Autonomous quantum error correction of Gottesman-Kitaev-Preskill states}},\ }\bibfield  {journal} {\bibinfo  {journal} {arXiv}\ }\href {https://doi.org/10.48550/arXiv.2310.11400} {10.48550/arXiv.2310.11400} (\bibinfo {year} {2023}),\ \Eprint {https://arxiv.org/abs/2310.11400} {2310.11400} \BibitemShut {NoStop}%
\bibitem [{\citenamefont {Puri}\ \emph {et~al.}(2020)\citenamefont {Puri}, \citenamefont {St-Jean}, \citenamefont {Gross}, \citenamefont {Grimm}, \citenamefont {Frattini}, \citenamefont {Iyer}, \citenamefont {Krishna}, \citenamefont {Touzard}, \citenamefont {Jiang}, \citenamefont {Blais}, \citenamefont {Flammia},\ and\ \citenamefont {Girvin}}]{Puri2020Aug}%
  \BibitemOpen
  \bibfield  {author} {\bibinfo {author} {\bibfnamefont {S.}~\bibnamefont {Puri}}, \bibinfo {author} {\bibfnamefont {L.}~\bibnamefont {St-Jean}}, \bibinfo {author} {\bibfnamefont {J.~A.}\ \bibnamefont {Gross}}, \bibinfo {author} {\bibfnamefont {A.}~\bibnamefont {Grimm}}, \bibinfo {author} {\bibfnamefont {N.~E.}\ \bibnamefont {Frattini}}, \bibinfo {author} {\bibfnamefont {P.~S.}\ \bibnamefont {Iyer}}, \bibinfo {author} {\bibfnamefont {A.}~\bibnamefont {Krishna}}, \bibinfo {author} {\bibfnamefont {S.}~\bibnamefont {Touzard}}, \bibinfo {author} {\bibfnamefont {L.}~\bibnamefont {Jiang}}, \bibinfo {author} {\bibfnamefont {A.}~\bibnamefont {Blais}}, \bibinfo {author} {\bibfnamefont {S.~T.}\ \bibnamefont {Flammia}},\ and\ \bibinfo {author} {\bibfnamefont {S.~M.}\ \bibnamefont {Girvin}},\ }\bibfield  {title} {\bibinfo {title} {{Bias-preserving gates with stabilized cat qubits}},\ }\bibfield  {journal} {\bibinfo  {journal} {Sci. Adv.}\ }\textbf {\bibinfo {volume} {6}},\ \href {https://doi.org/10.1126/sciadv.aay5901}
  {10.1126/sciadv.aay5901} (\bibinfo {year} {2020})\BibitemShut {NoStop}%
\bibitem [{\citenamefont {Kubica}\ and\ \citenamefont {Vasmer}(2022)}]{Kubica2022}%
  \BibitemOpen
  \bibfield  {author} {\bibinfo {author} {\bibfnamefont {A.}~\bibnamefont {Kubica}}\ and\ \bibinfo {author} {\bibfnamefont {M.}~\bibnamefont {Vasmer}},\ }\bibfield  {title} {\bibinfo {title} {Single-shot quantum error correction with the three-dimensional subsystem toric code},\ }\href {https://doi.org/10.1038/s41467-022-33923-4} {\bibfield  {journal} {\bibinfo  {journal} {Nature Communications}\ }\textbf {\bibinfo {volume} {13}},\ \bibinfo {pages} {6272} (\bibinfo {year} {2022})},\ \bibinfo {note} {introduces the 3D subsystem toric code with single-shot error correction}\BibitemShut {NoStop}%
\bibitem [{\citenamefont {Krinner}\ \emph {et~al.}(2022)\citenamefont {Krinner}, \citenamefont {Lacroix}, \citenamefont {Remm}, \citenamefont {Di~Paolo}, \citenamefont {Genois}, \citenamefont {Leroux}, \citenamefont {Hellings}, \citenamefont {Lazar}, \citenamefont {Swiadek}, \citenamefont {Herrmann}, \citenamefont {Norris}, \citenamefont {Andersen}, \citenamefont {M{\ifmmode\ddot{u}\else\"{u}\fi}ller}, \citenamefont {Blais}, \citenamefont {Eichler},\ and\ \citenamefont {Wallraff}}]{Krinner2022May}%
  \BibitemOpen
  \bibfield  {author} {\bibinfo {author} {\bibfnamefont {S.}~\bibnamefont {Krinner}}, \bibinfo {author} {\bibfnamefont {N.}~\bibnamefont {Lacroix}}, \bibinfo {author} {\bibfnamefont {A.}~\bibnamefont {Remm}}, \bibinfo {author} {\bibfnamefont {A.}~\bibnamefont {Di~Paolo}}, \bibinfo {author} {\bibfnamefont {E.}~\bibnamefont {Genois}}, \bibinfo {author} {\bibfnamefont {C.}~\bibnamefont {Leroux}}, \bibinfo {author} {\bibfnamefont {C.}~\bibnamefont {Hellings}}, \bibinfo {author} {\bibfnamefont {S.}~\bibnamefont {Lazar}}, \bibinfo {author} {\bibfnamefont {F.}~\bibnamefont {Swiadek}}, \bibinfo {author} {\bibfnamefont {J.}~\bibnamefont {Herrmann}}, \bibinfo {author} {\bibfnamefont {G.~J.}\ \bibnamefont {Norris}}, \bibinfo {author} {\bibfnamefont {C.~K.}\ \bibnamefont {Andersen}}, \bibinfo {author} {\bibfnamefont {M.}~\bibnamefont {M{\ifmmode\ddot{u}\else\"{u}\fi}ller}}, \bibinfo {author} {\bibfnamefont {A.}~\bibnamefont {Blais}}, \bibinfo {author} {\bibfnamefont {C.}~\bibnamefont {Eichler}},\ and\ \bibinfo {author}
  {\bibfnamefont {A.}~\bibnamefont {Wallraff}},\ }\bibfield  {title} {\bibinfo {title} {{Realizing repeated quantum error correction in a distance-three surface code}},\ }\href {https://doi.org/10.1038/s41586-022-04566-8} {\bibfield  {journal} {\bibinfo  {journal} {Nature}\ }\textbf {\bibinfo {volume} {605}},\ \bibinfo {pages} {669} (\bibinfo {year} {2022})}\BibitemShut {NoStop}%
\bibitem [{\citenamefont {Ginsberg}\ and\ \citenamefont {Patel}(2025)}]{ginsberg2025quantumerrordetectionearly}%
  \BibitemOpen
  \bibfield  {author} {\bibinfo {author} {\bibfnamefont {T.}~\bibnamefont {Ginsberg}}\ and\ \bibinfo {author} {\bibfnamefont {V.}~\bibnamefont {Patel}},\ }\href {https://arxiv.org/abs/2503.10790} {\bibinfo {title} {Quantum error detection for early term fault-tolerant quantum algorithms}} (\bibinfo {year} {2025}),\ \Eprint {https://arxiv.org/abs/2503.10790} {arXiv:2503.10790 [quant-ph]} \BibitemShut {NoStop}%
\bibitem [{\citenamefont {Torlai}\ and\ \citenamefont {Melko}(2017)}]{Torlai2017}%
  \BibitemOpen
  \bibfield  {author} {\bibinfo {author} {\bibfnamefont {G.}~\bibnamefont {Torlai}}\ and\ \bibinfo {author} {\bibfnamefont {R.~G.}\ \bibnamefont {Melko}},\ }\bibfield  {title} {\bibinfo {title} {Neural decoder for topological codes},\ }\href {https://doi.org/10.1103/PhysRevLett.119.030501} {\bibfield  {journal} {\bibinfo  {journal} {Physical Review Letters}\ }\textbf {\bibinfo {volume} {119}},\ \bibinfo {pages} {030501} (\bibinfo {year} {2017})},\ \Eprint {https://arxiv.org/abs/1610.04238} {arXiv:1610.04238 [quant-ph]} \BibitemShut {NoStop}%
\bibitem [{\citenamefont {Chamberland}\ and\ \citenamefont {Beverland}(2018)}]{Chamberland2018}%
  \BibitemOpen
  \bibfield  {author} {\bibinfo {author} {\bibfnamefont {C.}~\bibnamefont {Chamberland}}\ and\ \bibinfo {author} {\bibfnamefont {M.~E.}\ \bibnamefont {Beverland}},\ }\bibfield  {title} {\bibinfo {title} {Flag fault-tolerant error correction with arbitrary distance codes},\ }\href {https://doi.org/10.22331/q-2018-02-08-53} {\bibfield  {journal} {\bibinfo  {journal} {Quantum}\ }\textbf {\bibinfo {volume} {2}},\ \bibinfo {pages} {53} (\bibinfo {year} {2018})},\ \Eprint {https://arxiv.org/abs/1708.02246} {arXiv:1708.02246 [quant-ph]} \BibitemShut {NoStop}%
\bibitem [{\citenamefont {Lowe}\ \emph {et~al.}(2021)\citenamefont {Lowe}, \citenamefont {Gordon}, \citenamefont {Czarnik}, \citenamefont {Arrasmith}, \citenamefont {Coles},\ and\ \citenamefont {Cincio}}]{PhysRevResearch.3.033098}%
  \BibitemOpen
  \bibfield  {author} {\bibinfo {author} {\bibfnamefont {A.}~\bibnamefont {Lowe}}, \bibinfo {author} {\bibfnamefont {M.~H.}\ \bibnamefont {Gordon}}, \bibinfo {author} {\bibfnamefont {P.}~\bibnamefont {Czarnik}}, \bibinfo {author} {\bibfnamefont {A.}~\bibnamefont {Arrasmith}}, \bibinfo {author} {\bibfnamefont {P.~J.}\ \bibnamefont {Coles}},\ and\ \bibinfo {author} {\bibfnamefont {L.}~\bibnamefont {Cincio}},\ }\bibfield  {title} {\bibinfo {title} {Unified approach to data-driven quantum error mitigation},\ }\href {https://doi.org/10.1103/PhysRevResearch.3.033098} {\bibfield  {journal} {\bibinfo  {journal} {Phys. Rev. Res.}\ }\textbf {\bibinfo {volume} {3}},\ \bibinfo {pages} {033098} (\bibinfo {year} {2021})}\BibitemShut {NoStop}%
\bibitem [{\citenamefont {Raymond}\ \emph {et~al.}(2025)\citenamefont {Raymond}, \citenamefont {Amin}, \citenamefont {King}, \citenamefont {Harris}, \citenamefont {Bernoudy}, \citenamefont {Berkley}, \citenamefont {Boothby}, \citenamefont {Smirnov}, \citenamefont {Altomare}, \citenamefont {Babcock}, \citenamefont {Baron}, \citenamefont {Connor}, \citenamefont {Dehn}, \citenamefont {Enderud}, \citenamefont {Hoskinson}, \citenamefont {Huang}, \citenamefont {Johnson}, \citenamefont {Ladizinsky}, \citenamefont {Lanting}, \citenamefont {MacDonald}, \citenamefont {Marsden}, \citenamefont {Molavi}, \citenamefont {Oh}, \citenamefont {Poulin-Lamarre}, \citenamefont {Ramp}, \citenamefont {Rich}, \citenamefont {Clavera}, \citenamefont {Tsai}, \citenamefont {Volkmann}, \citenamefont {Whittaker}, \citenamefont {Yao}, \citenamefont {Heinsdorf}, \citenamefont {Kaushal}, \citenamefont {Nocera}, \citenamefont {Franz},\ and\ \citenamefont {Dziarmaga}}]{Raymond2025Mar}%
  \BibitemOpen
  \bibfield  {author} {\bibinfo {author} {\bibfnamefont {J.}~\bibnamefont {Raymond}}, \bibinfo {author} {\bibfnamefont {M.~H.}\ \bibnamefont {Amin}}, \bibinfo {author} {\bibfnamefont {A.~D.}\ \bibnamefont {King}}, \bibinfo {author} {\bibfnamefont {R.}~\bibnamefont {Harris}}, \bibinfo {author} {\bibfnamefont {W.}~\bibnamefont {Bernoudy}}, \bibinfo {author} {\bibfnamefont {A.~J.}\ \bibnamefont {Berkley}}, \bibinfo {author} {\bibfnamefont {K.}~\bibnamefont {Boothby}}, \bibinfo {author} {\bibfnamefont {A.}~\bibnamefont {Smirnov}}, \bibinfo {author} {\bibfnamefont {F.}~\bibnamefont {Altomare}}, \bibinfo {author} {\bibfnamefont {M.}~\bibnamefont {Babcock}}, \bibinfo {author} {\bibfnamefont {C.}~\bibnamefont {Baron}}, \bibinfo {author} {\bibfnamefont {J.}~\bibnamefont {Connor}}, \bibinfo {author} {\bibfnamefont {M.~H.}\ \bibnamefont {Dehn}}, \bibinfo {author} {\bibfnamefont {C.}~\bibnamefont {Enderud}}, \bibinfo {author} {\bibfnamefont {E.}~\bibnamefont {Hoskinson}}, \bibinfo {author} {\bibfnamefont
  {S.}~\bibnamefont {Huang}}, \bibinfo {author} {\bibfnamefont {M.~W.}\ \bibnamefont {Johnson}}, \bibinfo {author} {\bibfnamefont {E.}~\bibnamefont {Ladizinsky}}, \bibinfo {author} {\bibfnamefont {T.}~\bibnamefont {Lanting}}, \bibinfo {author} {\bibfnamefont {A.~J.~R.}\ \bibnamefont {MacDonald}}, \bibinfo {author} {\bibfnamefont {G.}~\bibnamefont {Marsden}}, \bibinfo {author} {\bibfnamefont {R.}~\bibnamefont {Molavi}}, \bibinfo {author} {\bibfnamefont {T.}~\bibnamefont {Oh}}, \bibinfo {author} {\bibfnamefont {G.}~\bibnamefont {Poulin-Lamarre}}, \bibinfo {author} {\bibfnamefont {H.}~\bibnamefont {Ramp}}, \bibinfo {author} {\bibfnamefont {C.}~\bibnamefont {Rich}}, \bibinfo {author} {\bibfnamefont {B.~T.}\ \bibnamefont {Clavera}}, \bibinfo {author} {\bibfnamefont {N.}~\bibnamefont {Tsai}}, \bibinfo {author} {\bibfnamefont {M.}~\bibnamefont {Volkmann}}, \bibinfo {author} {\bibfnamefont {J.~D.}\ \bibnamefont {Whittaker}}, \bibinfo {author} {\bibfnamefont {J.}~\bibnamefont {Yao}}, \bibinfo {author} {\bibfnamefont
  {N.}~\bibnamefont {Heinsdorf}}, \bibinfo {author} {\bibfnamefont {N.}~\bibnamefont {Kaushal}}, \bibinfo {author} {\bibfnamefont {A.}~\bibnamefont {Nocera}}, \bibinfo {author} {\bibfnamefont {M.}~\bibnamefont {Franz}},\ and\ \bibinfo {author} {\bibfnamefont {J.}~\bibnamefont {Dziarmaga}},\ }\bibfield  {title} {\bibinfo {title} {{Quantum error mitigation in quantum annealing}},\ }\href {https://doi.org/10.1038/s41534-025-00977-3} {\bibfield  {journal} {\bibinfo  {journal} {npj Quantum Inf.}\ }\textbf {\bibinfo {volume} {11}},\ \bibinfo {pages} {1} (\bibinfo {year} {2025})}\BibitemShut {NoStop}%
\bibitem [{\citenamefont {Ijaz}\ \emph {et~al.}(2025)\citenamefont {Ijaz}, \citenamefont {Alderete}, \citenamefont {Sauvage}, \citenamefont {Cincio}, \citenamefont {Cerezo},\ and\ \citenamefont {Goh}}]{IJAZ2025100042}%
  \BibitemOpen
  \bibfield  {author} {\bibinfo {author} {\bibfnamefont {A.}~\bibnamefont {Ijaz}}, \bibinfo {author} {\bibfnamefont {C.~H.}\ \bibnamefont {Alderete}}, \bibinfo {author} {\bibfnamefont {F.}~\bibnamefont {Sauvage}}, \bibinfo {author} {\bibfnamefont {L.}~\bibnamefont {Cincio}}, \bibinfo {author} {\bibfnamefont {M.}~\bibnamefont {Cerezo}},\ and\ \bibinfo {author} {\bibfnamefont {M.~L.}\ \bibnamefont {Goh}},\ }\bibfield  {title} {\bibinfo {title} {More buck-per-shot: Why learning trumps mitigation in noisy quantum sensing},\ }\href {https://doi.org/https://doi.org/10.1016/j.mtquan.2025.100042} {\bibfield  {journal} {\bibinfo  {journal} {Materials Today Quantum}\ }\textbf {\bibinfo {volume} {6}},\ \bibinfo {pages} {100042} (\bibinfo {year} {2025})}\BibitemShut {NoStop}%
\bibitem [{\citenamefont {Wallman}\ and\ \citenamefont {Emerson}(2016)}]{PhysRevA.94.052325}%
  \BibitemOpen
  \bibfield  {author} {\bibinfo {author} {\bibfnamefont {J.~J.}\ \bibnamefont {Wallman}}\ and\ \bibinfo {author} {\bibfnamefont {J.}~\bibnamefont {Emerson}},\ }\bibfield  {title} {\bibinfo {title} {Noise tailoring for scalable quantum computation via randomized compiling},\ }\href {https://doi.org/10.1103/PhysRevA.94.052325} {\bibfield  {journal} {\bibinfo  {journal} {Phys. Rev. A}\ }\textbf {\bibinfo {volume} {94}},\ \bibinfo {pages} {052325} (\bibinfo {year} {2016})}\BibitemShut {NoStop}%
\bibitem [{\citenamefont {Hashim}\ \emph {et~al.}(2021)\citenamefont {Hashim}, \citenamefont {Naik}, \citenamefont {Morvan}, \citenamefont {Ville}, \citenamefont {Mitchell}, \citenamefont {Kreikebaum}, \citenamefont {Davis}, \citenamefont {Smith}, \citenamefont {Iancu}, \citenamefont {O'Brien}, \citenamefont {Hincks}, \citenamefont {Wallman}, \citenamefont {Emerson},\ and\ \citenamefont {Siddiqi}}]{PhysRevX.11.041039}%
  \BibitemOpen
  \bibfield  {author} {\bibinfo {author} {\bibfnamefont {A.}~\bibnamefont {Hashim}}, \bibinfo {author} {\bibfnamefont {R.~K.}\ \bibnamefont {Naik}}, \bibinfo {author} {\bibfnamefont {A.}~\bibnamefont {Morvan}}, \bibinfo {author} {\bibfnamefont {J.-L.}\ \bibnamefont {Ville}}, \bibinfo {author} {\bibfnamefont {B.}~\bibnamefont {Mitchell}}, \bibinfo {author} {\bibfnamefont {J.~M.}\ \bibnamefont {Kreikebaum}}, \bibinfo {author} {\bibfnamefont {M.}~\bibnamefont {Davis}}, \bibinfo {author} {\bibfnamefont {E.}~\bibnamefont {Smith}}, \bibinfo {author} {\bibfnamefont {C.}~\bibnamefont {Iancu}}, \bibinfo {author} {\bibfnamefont {K.~P.}\ \bibnamefont {O'Brien}}, \bibinfo {author} {\bibfnamefont {I.}~\bibnamefont {Hincks}}, \bibinfo {author} {\bibfnamefont {J.~J.}\ \bibnamefont {Wallman}}, \bibinfo {author} {\bibfnamefont {J.}~\bibnamefont {Emerson}},\ and\ \bibinfo {author} {\bibfnamefont {I.}~\bibnamefont {Siddiqi}},\ }\bibfield  {title} {\bibinfo {title} {Randomized compiling for scalable quantum computing on a noisy
  superconducting quantum processor},\ }\href {https://doi.org/10.1103/PhysRevX.11.041039} {\bibfield  {journal} {\bibinfo  {journal} {Phys. Rev. X}\ }\textbf {\bibinfo {volume} {11}},\ \bibinfo {pages} {041039} (\bibinfo {year} {2021})}\BibitemShut {NoStop}%
\bibitem [{\citenamefont {Winick}\ \emph {et~al.}(2022)\citenamefont {Winick}, \citenamefont {Wallman}, \citenamefont {Dahlen}, \citenamefont {Hincks}, \citenamefont {Ospadov},\ and\ \citenamefont {Emerson}}]{winick2022conceptsconditionserrorsuppression}%
  \BibitemOpen
  \bibfield  {author} {\bibinfo {author} {\bibfnamefont {A.}~\bibnamefont {Winick}}, \bibinfo {author} {\bibfnamefont {J.~J.}\ \bibnamefont {Wallman}}, \bibinfo {author} {\bibfnamefont {D.}~\bibnamefont {Dahlen}}, \bibinfo {author} {\bibfnamefont {I.}~\bibnamefont {Hincks}}, \bibinfo {author} {\bibfnamefont {E.}~\bibnamefont {Ospadov}},\ and\ \bibinfo {author} {\bibfnamefont {J.}~\bibnamefont {Emerson}},\ }\href {https://arxiv.org/abs/2212.07500} {\bibinfo {title} {Concepts and conditions for error suppression through randomized compiling}} (\bibinfo {year} {2022}),\ \Eprint {https://arxiv.org/abs/2212.07500} {arXiv:2212.07500 [quant-ph]} \BibitemShut {NoStop}%
\bibitem [{\citenamefont {Beale}\ and\ \citenamefont {Wallman}(2023)}]{beale2023subsystemmeasurements}%
  \BibitemOpen
  \bibfield  {author} {\bibinfo {author} {\bibfnamefont {S.~J.}\ \bibnamefont {Beale}}\ and\ \bibinfo {author} {\bibfnamefont {J.~J.}\ \bibnamefont {Wallman}},\ }\href {https://arxiv.org/abs/2304.06599} {\bibinfo {title} {Randomized compiling for subsystem measurements}} (\bibinfo {year} {2023}),\ \Eprint {https://arxiv.org/abs/2304.06599} {arXiv:2304.06599 [quant-ph]} \BibitemShut {NoStop}%
\bibitem [{\citenamefont {Miguel-Ramiro}\ \emph {et~al.}(2023)\citenamefont {Miguel-Ramiro}, \citenamefont {Shi}, \citenamefont {Dellantonio}, \citenamefont {Chan}, \citenamefont {Muschik},\ and\ \citenamefont {Dür}}]{Miguel_Ramiro_2023}%
  \BibitemOpen
  \bibfield  {author} {\bibinfo {author} {\bibfnamefont {J.}~\bibnamefont {Miguel-Ramiro}}, \bibinfo {author} {\bibfnamefont {Z.}~\bibnamefont {Shi}}, \bibinfo {author} {\bibfnamefont {L.}~\bibnamefont {Dellantonio}}, \bibinfo {author} {\bibfnamefont {A.}~\bibnamefont {Chan}}, \bibinfo {author} {\bibfnamefont {C.~A.}\ \bibnamefont {Muschik}},\ and\ \bibinfo {author} {\bibfnamefont {W.}~\bibnamefont {Dür}},\ }\bibfield  {title} {\bibinfo {title} {Superposed quantum error mitigation},\ }\bibfield  {journal} {\bibinfo  {journal} {Physical Review Letters}\ }\textbf {\bibinfo {volume} {131}},\ \href {https://doi.org/10.1103/physrevlett.131.230601} {10.1103/physrevlett.131.230601} (\bibinfo {year} {2023})\BibitemShut {NoStop}%
\bibitem [{\citenamefont {Ferracin}\ \emph {et~al.}(2024)\citenamefont {Ferracin}, \citenamefont {Hashim}, \citenamefont {Ville}, \citenamefont {Naik}, \citenamefont {Carignan-Dugas}, \citenamefont {Qassim}, \citenamefont {Morvan}, \citenamefont {Santiago}, \citenamefont {Siddiqi},\ and\ \citenamefont {Wallman}}]{Ferracin2024efficiently}%
  \BibitemOpen
  \bibfield  {author} {\bibinfo {author} {\bibfnamefont {S.}~\bibnamefont {Ferracin}}, \bibinfo {author} {\bibfnamefont {A.}~\bibnamefont {Hashim}}, \bibinfo {author} {\bibfnamefont {J.-L.}\ \bibnamefont {Ville}}, \bibinfo {author} {\bibfnamefont {R.}~\bibnamefont {Naik}}, \bibinfo {author} {\bibfnamefont {A.}~\bibnamefont {Carignan-Dugas}}, \bibinfo {author} {\bibfnamefont {H.}~\bibnamefont {Qassim}}, \bibinfo {author} {\bibfnamefont {A.}~\bibnamefont {Morvan}}, \bibinfo {author} {\bibfnamefont {D.~I.}\ \bibnamefont {Santiago}}, \bibinfo {author} {\bibfnamefont {I.}~\bibnamefont {Siddiqi}},\ and\ \bibinfo {author} {\bibfnamefont {J.~J.}\ \bibnamefont {Wallman}},\ }\bibfield  {title} {\bibinfo {title} {Efficiently improving the performance of noisy quantum computers},\ }\href {https://doi.org/10.22331/q-2024-07-15-1410} {\bibfield  {journal} {\bibinfo  {journal} {{Quantum}}\ }\textbf {\bibinfo {volume} {8}},\ \bibinfo {pages} {1410} (\bibinfo {year} {2024})}\BibitemShut {NoStop}%
\bibitem [{\citenamefont {Saxena}\ and\ \citenamefont {Kyaw}(2024)}]{saxena2024errormitigationrestrictedevolution}%
  \BibitemOpen
  \bibfield  {author} {\bibinfo {author} {\bibfnamefont {G.}~\bibnamefont {Saxena}}\ and\ \bibinfo {author} {\bibfnamefont {T.~H.}\ \bibnamefont {Kyaw}},\ }\href {https://arxiv.org/abs/2409.06636} {\bibinfo {title} {Error mitigation by restricted evolution}} (\bibinfo {year} {2024}),\ \Eprint {https://arxiv.org/abs/2409.06636} {arXiv:2409.06636 [quant-ph]} \BibitemShut {NoStop}%
\bibitem [{\citenamefont {Díez-Valle}\ \emph {et~al.}(2025)\citenamefont {Díez-Valle}, \citenamefont {Saxena}, \citenamefont {Baker}, \citenamefont {Lee},\ and\ \citenamefont {Kyaw}}]{díezvalle_2025}%
  \BibitemOpen
  \bibfield  {author} {\bibinfo {author} {\bibfnamefont {P.}~\bibnamefont {Díez-Valle}}, \bibinfo {author} {\bibfnamefont {G.}~\bibnamefont {Saxena}}, \bibinfo {author} {\bibfnamefont {J.~S.}\ \bibnamefont {Baker}}, \bibinfo {author} {\bibfnamefont {J.-H.}\ \bibnamefont {Lee}},\ and\ \bibinfo {author} {\bibfnamefont {T.~H.}\ \bibnamefont {Kyaw}},\ }\href {https://arxiv.org/abs/2505.07977} {\bibinfo {title} {Physically motivated extrapolation for quantum error mitigation}} (\bibinfo {year} {2025}),\ \Eprint {https://arxiv.org/abs/2505.07977} {arXiv:2505.07977 [quant-ph]} \BibitemShut {NoStop}%
\bibitem [{\citenamefont {Bennewitz}\ \emph {et~al.}(2022)\citenamefont {Bennewitz}, \citenamefont {Hopfmueller}, \citenamefont {Kulchytskyy}, \citenamefont {Carrasquilla},\ and\ \citenamefont {Ronagh}}]{Bennewitz_2022}%
  \BibitemOpen
  \bibfield  {author} {\bibinfo {author} {\bibfnamefont {E.~R.}\ \bibnamefont {Bennewitz}}, \bibinfo {author} {\bibfnamefont {F.}~\bibnamefont {Hopfmueller}}, \bibinfo {author} {\bibfnamefont {B.}~\bibnamefont {Kulchytskyy}}, \bibinfo {author} {\bibfnamefont {J.}~\bibnamefont {Carrasquilla}},\ and\ \bibinfo {author} {\bibfnamefont {P.}~\bibnamefont {Ronagh}},\ }\bibfield  {title} {\bibinfo {title} {Neural error mitigation of near-term quantum simulations},\ }\href {https://doi.org/10.1038/s42256-022-00509-0} {\bibfield  {journal} {\bibinfo  {journal} {Nature Machine Intelligence}\ }\textbf {\bibinfo {volume} {4}},\ \bibinfo {pages} {618–624} (\bibinfo {year} {2022})}\BibitemShut {NoStop}%
\bibitem [{\citenamefont {Liao}\ \emph {et~al.}(2025)\citenamefont {Liao}, \citenamefont {Zhu}, \citenamefont {Chiribella},\ and\ \citenamefont {Yang}}]{Liao2025Jan}%
  \BibitemOpen
  \bibfield  {author} {\bibinfo {author} {\bibfnamefont {M.}~\bibnamefont {Liao}}, \bibinfo {author} {\bibfnamefont {Y.}~\bibnamefont {Zhu}}, \bibinfo {author} {\bibfnamefont {G.}~\bibnamefont {Chiribella}},\ and\ \bibinfo {author} {\bibfnamefont {Y.}~\bibnamefont {Yang}},\ }\bibfield  {title} {\bibinfo {title} {{Noise-agnostic quantum error mitigation with data augmented neural models}},\ }\href {https://doi.org/10.1038/s41534-025-00960-y} {\bibfield  {journal} {\bibinfo  {journal} {npj Quantum Inf.}\ }\textbf {\bibinfo {volume} {11}},\ \bibinfo {pages} {1} (\bibinfo {year} {2025})}\BibitemShut {NoStop}%
\bibitem [{\citenamefont {Su}\ \emph {et~al.}(2021)\citenamefont {Su}, \citenamefont {Israel}, \citenamefont {Sharma}, \citenamefont {Qi}, \citenamefont {Dhand},\ and\ \citenamefont {Br{\'{a}}dler}}]{Su2021errormitigationnear}%
  \BibitemOpen
  \bibfield  {author} {\bibinfo {author} {\bibfnamefont {D.}~\bibnamefont {Su}}, \bibinfo {author} {\bibfnamefont {R.}~\bibnamefont {Israel}}, \bibinfo {author} {\bibfnamefont {K.}~\bibnamefont {Sharma}}, \bibinfo {author} {\bibfnamefont {H.}~\bibnamefont {Qi}}, \bibinfo {author} {\bibfnamefont {I.}~\bibnamefont {Dhand}},\ and\ \bibinfo {author} {\bibfnamefont {K.}~\bibnamefont {Br{\'{a}}dler}},\ }\bibfield  {title} {\bibinfo {title} {Error mitigation on a near-term quantum photonic device},\ }\href {https://doi.org/10.22331/q-2021-05-04-452} {\bibfield  {journal} {\bibinfo  {journal} {{Quantum}}\ }\textbf {\bibinfo {volume} {5}},\ \bibinfo {pages} {452} (\bibinfo {year} {2021})}\BibitemShut {NoStop}%
\bibitem [{\citenamefont {Prasad}\ \emph {et~al.}(2024)\citenamefont {Prasad}, \citenamefont {Cheng}, \citenamefont {Fekl},\ and\ \citenamefont {Jacobsen}}]{D3CP03523A}%
  \BibitemOpen
  \bibfield  {author} {\bibinfo {author} {\bibfnamefont {V.~K.}\ \bibnamefont {Prasad}}, \bibinfo {author} {\bibfnamefont {F.}~\bibnamefont {Cheng}}, \bibinfo {author} {\bibfnamefont {U.}~\bibnamefont {Fekl}},\ and\ \bibinfo {author} {\bibfnamefont {H.-A.}\ \bibnamefont {Jacobsen}},\ }\bibfield  {title} {\bibinfo {title} {Applications of noisy quantum computing and quantum error mitigation to “adamantaneland”: a benchmarking study for quantum chemistry},\ }\href {https://doi.org/10.1039/D3CP03523A} {\bibfield  {journal} {\bibinfo  {journal} {Phys. Chem. Chem. Phys.}\ }\textbf {\bibinfo {volume} {26}},\ \bibinfo {pages} {4071} (\bibinfo {year} {2024})}\BibitemShut {NoStop}%
\bibitem [{\citenamefont {Hagge}\ and\ \citenamefont {Wiebe}(2023)}]{hagge2023errormitigationerrordetection}%
  \BibitemOpen
  \bibfield  {author} {\bibinfo {author} {\bibfnamefont {T.}~\bibnamefont {Hagge}}\ and\ \bibinfo {author} {\bibfnamefont {N.}~\bibnamefont {Wiebe}},\ }\href {https://arxiv.org/abs/2309.11673} {\bibinfo {title} {Error mitigation via error detection using generalized superfast encodings}} (\bibinfo {year} {2023}),\ \Eprint {https://arxiv.org/abs/2309.11673} {arXiv:2309.11673 [quant-ph]} \BibitemShut {NoStop}%
\bibitem [{\citenamefont {Lin}\ \emph {et~al.}(2021)\citenamefont {Lin}, \citenamefont {Wallman}, \citenamefont {Hincks},\ and\ \citenamefont {Laflamme}}]{PhysRevResearch.3.033285}%
  \BibitemOpen
  \bibfield  {author} {\bibinfo {author} {\bibfnamefont {J.}~\bibnamefont {Lin}}, \bibinfo {author} {\bibfnamefont {J.~J.}\ \bibnamefont {Wallman}}, \bibinfo {author} {\bibfnamefont {I.}~\bibnamefont {Hincks}},\ and\ \bibinfo {author} {\bibfnamefont {R.}~\bibnamefont {Laflamme}},\ }\bibfield  {title} {\bibinfo {title} {Independent state and measurement characterization for quantum computers},\ }\href {https://doi.org/10.1103/PhysRevResearch.3.033285} {\bibfield  {journal} {\bibinfo  {journal} {Phys. Rev. Res.}\ }\textbf {\bibinfo {volume} {3}},\ \bibinfo {pages} {033285} (\bibinfo {year} {2021})}\BibitemShut {NoStop}%
\bibitem [{\citenamefont {LaRose}\ \emph {et~al.}(2022)\citenamefont {LaRose}, \citenamefont {Mari}, \citenamefont {Kaiser}, \citenamefont {Karalekas}, \citenamefont {Alves}, \citenamefont {Czarnik}, \citenamefont {El~Mandouh}, \citenamefont {Gordon}, \citenamefont {Hindy}, \citenamefont {Robertson}, \citenamefont {Thakre}, \citenamefont {Wahl}, \citenamefont {Samuel}, \citenamefont {Mistri}, \citenamefont {Tremblay}, \citenamefont {Gardner}, \citenamefont {Stemen}, \citenamefont {Shammah},\ and\ \citenamefont {Zeng}}]{LaRose2022mitiqsoftware}%
  \BibitemOpen
  \bibfield  {author} {\bibinfo {author} {\bibfnamefont {R.}~\bibnamefont {LaRose}}, \bibinfo {author} {\bibfnamefont {A.}~\bibnamefont {Mari}}, \bibinfo {author} {\bibfnamefont {S.}~\bibnamefont {Kaiser}}, \bibinfo {author} {\bibfnamefont {P.~J.}\ \bibnamefont {Karalekas}}, \bibinfo {author} {\bibfnamefont {A.~A.}\ \bibnamefont {Alves}}, \bibinfo {author} {\bibfnamefont {P.}~\bibnamefont {Czarnik}}, \bibinfo {author} {\bibfnamefont {M.}~\bibnamefont {El~Mandouh}}, \bibinfo {author} {\bibfnamefont {M.~H.}\ \bibnamefont {Gordon}}, \bibinfo {author} {\bibfnamefont {Y.}~\bibnamefont {Hindy}}, \bibinfo {author} {\bibfnamefont {A.}~\bibnamefont {Robertson}}, \bibinfo {author} {\bibfnamefont {P.}~\bibnamefont {Thakre}}, \bibinfo {author} {\bibfnamefont {M.}~\bibnamefont {Wahl}}, \bibinfo {author} {\bibfnamefont {D.}~\bibnamefont {Samuel}}, \bibinfo {author} {\bibfnamefont {R.}~\bibnamefont {Mistri}}, \bibinfo {author} {\bibfnamefont {M.}~\bibnamefont {Tremblay}}, \bibinfo {author} {\bibfnamefont {N.}~\bibnamefont
  {Gardner}}, \bibinfo {author} {\bibfnamefont {N.~T.}\ \bibnamefont {Stemen}}, \bibinfo {author} {\bibfnamefont {N.}~\bibnamefont {Shammah}},\ and\ \bibinfo {author} {\bibfnamefont {W.~J.}\ \bibnamefont {Zeng}},\ }\bibfield  {title} {\bibinfo {title} {Mitiq: {A} software package for error mitigation on noisy quantum computers},\ }\href {https://doi.org/10.22331/q-2022-08-11-774} {\bibfield  {journal} {\bibinfo  {journal} {{Quantum}}\ }\textbf {\bibinfo {volume} {6}},\ \bibinfo {pages} {774} (\bibinfo {year} {2022})}\BibitemShut {NoStop}%
\bibitem [{\citenamefont {Lee}\ \emph {et~al.}(2021)\citenamefont {Lee}, \citenamefont {Berry}, \citenamefont {Gidney}, \citenamefont {Huggins}, \citenamefont {McClean}, \citenamefont {Wiebe},\ and\ \citenamefont {Babbush}}]{PRXQuantum.2.030305}%
  \BibitemOpen
  \bibfield  {author} {\bibinfo {author} {\bibfnamefont {J.}~\bibnamefont {Lee}}, \bibinfo {author} {\bibfnamefont {D.~W.}\ \bibnamefont {Berry}}, \bibinfo {author} {\bibfnamefont {C.}~\bibnamefont {Gidney}}, \bibinfo {author} {\bibfnamefont {W.~J.}\ \bibnamefont {Huggins}}, \bibinfo {author} {\bibfnamefont {J.~R.}\ \bibnamefont {McClean}}, \bibinfo {author} {\bibfnamefont {N.}~\bibnamefont {Wiebe}},\ and\ \bibinfo {author} {\bibfnamefont {R.}~\bibnamefont {Babbush}},\ }\bibfield  {title} {\bibinfo {title} {Even more efficient quantum computations of chemistry through tensor hypercontraction},\ }\href {https://doi.org/10.1103/PRXQuantum.2.030305} {\bibfield  {journal} {\bibinfo  {journal} {PRX Quantum}\ }\textbf {\bibinfo {volume} {2}},\ \bibinfo {pages} {030305} (\bibinfo {year} {2021})}\BibitemShut {NoStop}%
\bibitem [{\citenamefont {Suzuki}\ \emph {et~al.}(2022)\citenamefont {Suzuki}, \citenamefont {Endo}, \citenamefont {Fujii},\ and\ \citenamefont {Tokunaga}}]{PRXQuantum.3.010345}%
  \BibitemOpen
  \bibfield  {author} {\bibinfo {author} {\bibfnamefont {Y.}~\bibnamefont {Suzuki}}, \bibinfo {author} {\bibfnamefont {S.}~\bibnamefont {Endo}}, \bibinfo {author} {\bibfnamefont {K.}~\bibnamefont {Fujii}},\ and\ \bibinfo {author} {\bibfnamefont {Y.}~\bibnamefont {Tokunaga}},\ }\bibfield  {title} {\bibinfo {title} {Quantum error mitigation as a universal error reduction technique: Applications from the nisq to the fault-tolerant quantum computing eras},\ }\href {https://doi.org/10.1103/PRXQuantum.3.010345} {\bibfield  {journal} {\bibinfo  {journal} {PRX Quantum}\ }\textbf {\bibinfo {volume} {3}},\ \bibinfo {pages} {010345} (\bibinfo {year} {2022})}\BibitemShut {NoStop}%
\end{thebibliography}%
\end{document}